\documentclass[twoside,british]{iopart}
\usepackage[T1]{fontenc}
\usepackage[utf8]{inputenc}
\usepackage{geometry}
\usepackage{verbatim}
\usepackage{textcomp}
\usepackage{amssymb}
\usepackage{esint}
\usepackage{graphicx}

\makeatletter
\usepackage{iopams}
\usepackage{setstack}

\usepackage[numbers, sort&compress]{natbib}

\newcommand{\binom}[2]{\left(\begin{array}{c}#1\\#2\end{array}\right)}

\renewenvironment{cases}{\left\{\begin{array}{ll}}{\end{array}\right.}

\makeatother

\usepackage{babel}
\begin{document}

\global\long\def\Hil{\mathcal{H}}
\global\long\def\SL{SL(2,\mathbb{C})}
\global\long\def\T{\mathbb{T}}
\global\long\def\F{\,_{2}F_{1}}

\global\long\def\dd{\mathrm{d}}
\global\long\def\im{\dot{\imath}}
\global\long\def\fjm{f_{m}^{(j)}}
\global\long\def\fjmi{f_{m_{i}}^{(j_{i})}}
\global\long\def\fjmn{\fjm\left(\eta\right)}
\global\long\def\fjmni{\fjmi\left(\eta\right)}

\global\long\def\at#1#2{\left.#1\right|_{#2}}
\global\long\def\atzero#1{\at{#1}{\eta=0}}

\global\long\def\deta#1{\frac{\dd#1}{\dd\eta}}
\global\long\def\detadwa#1{\frac{\dd^{2}#1}{\dd\eta^{2}}}
\global\long\def\detazero#1{\atzero{\deta{#1}}}
\global\long\def\detadwazero#1{\atzero{\detadwa{#1}}}

\global\long\def\FF#1#2#3#4{\F\left(#1,\,#2;\,#3;\,#4\right)}

\global\long\def\ket#1{\left|#1\right\rangle }
\global\long\def\bra#1{\left\langle #1\right|}

\global\long\def\ketbra#1{\ket{#1}\bra{#1}}

\title{Asymptotic of Lorentzian Polyhedra Propagator}

\author{{{Jacek Puchta}}}

\address{{{Instytut Fizyki Teoretycznej, Uniwersytet Warszawski, ul.
Ho\.{z}a 69, 00-681 Warsaw, Poland}}}

\address{{{Centre de Physique Théorique de Luminy, Case 907, Luminy,
F-13288 Marseille, France}}}

\ead{{{jpa@fuw.edu.pl}}}
\begin{abstract}
A certain operator $\T=\int_{\SL}\dd g\, Y^{\dagger}gY$ can be found
in various Lorentzian EPRL calculations. The properties of this operator
has been studied here in large $j$ limit. The leading order of $\T$
is proportional to the identity operator.

Knowing the operator $\T$ one can renormalize spin-foam's edge self-energy by computing the amplitude of sum of a series of edges with increasing number of vertices and bubbles. This amplitude is calculated and is shown to be convergent.

Moreover some technical tools useful in Lorentzian Spin-Foam calculation
has been developed.
\end{abstract}





\section{Introduction and motivation}

Loop quantum gravity and its covariant version - Spin-Foam Models
- are promising candidates for a quantum theory of gravity \cite{LQGHistory,LQGBasics}.
Recent development has shown that within the Spin-Foam Models the
EPRL method of calculating the vertex amplitude \cite{EPRL,EucleadianEPRL,LorentzianEPRL,OperatorSF,SimplicityConstraint,L-S,NewLook}
are a way to recover the simplicity constraints of general relativity
in the classical limit \cite{SimplicityConstraint,CarloLectures}.
There are two main classes of the EPRL model: the euclidean EPRL \cite{EPRL,EucleadianEPRL}
- with the $SO(4)$ group as a gauge group of the theory, and the
Lorentzian EPRL \cite{EPRL,LorentzianEPRL} - with the universal covering
group of $SO(1,3)$, namely $\SL$, as a gauge group. In both versions
of the model the key role is played by so called EPRL map being the
map
\begin{equation}
Y\,:\,\bigoplus_{j}\Hil_{j}\to\bigoplus_{\rho}\Hil_{\rho}
\end{equation}
where $j$ are the spinlabels of representations of $SU(2)$ and $\Hil_{j}$
is the Hilbert space, on which this representation acts, while $\rho$
are the unitary representations of the group $G$ (being either $SO(4)$
or $\SL$) and $\Hil_{\rho}$ is the representation space respectively.
In the euclidean case the $Y$ map is given by certain combinations
of Clebsh-Gordon coefficients, and the $SO(4)$ representations can
be expressed in terms of Wigner matrices, whose properties are well
known. These makes the calculations relatively simple and several
examples has been studied in detail. Calculations in Lorentzian case
however involve matrix elements of unitary $\SL$ representations,
being combinations of hypergeometric functions \cite{Ruhl}, what
increase a lot difficulty of getting a precise result.

Nevertheless some attempts has been done in the Lorentzian model as
well. In the Dipole Cosmology model (DC) \cite{DC_Lorentzian} the
transition amplitude is written explicitly
\begin{eqnarray}
W(z) & = & \sum_{\left\{ j_{\ell}\right\} }\prod_{\ell=1}^{4}\left(2j_{\ell}+1\right)e^{-2t\hbar j_{\ell}(j_{\ell}+1)-\im\lambda v_{0}j_{\ell}^{\frac{3}{2}}-\im zj_{\ell}}\int_{\SL}\dd g\prod_{\ell=1}^{4}\bra{j_{\ell}}u_{\vec{n_{\ell}}}^{\dagger}Y^{\dagger}g\, Y\, u_{\vec{n'_{\ell}}}\ket{j_{\ell}}_{j_{\ell}}\nonumber \\
\  & = & \sum_{\left\{ j_{\ell}\right\} }\prod_{\ell=1}^{4}\left(2j_{\ell}+1\right)e^{-2t\hbar j_{\ell}(j_{\ell}+1)-\im\lambda v_{0}j_{\ell}^{\frac{3}{2}}-\im zj_{\ell}}\,\bra{\iota}\T\ket{\iota'}\\
\  & \  & \ \ \ {\rm with}\ \ \iota^{(')}:=\int_{SU(2)}\dd u\,\prod_{\ell=1}^{4}u\cdot u_{\vec{n^{(')}}}\ket{j_{\ell}}\nonumber 
\end{eqnarray}
The integral does not have to be known explicitly, the only thing
authors need is that it is bound by $N_{0}j_{0}^{-3}$ for large $j_{0}$.
Such behaviour of this integral is assumed per analogy to the euclidean
case \cite{DC1}, however it was not proven rigorously.

The same integral can be found in the calculation of radiative correction
to the spin-foam edge (also called the ''self energy'' of the edge)
coming from the ''melonic'' diagram \cite{Aldo}. In this paper it
was shown, that the divergent part coming from a certain bubble is
\begin{eqnarray}
W^{\Lambda} & \sim & \Lambda^{6(\mu-1)}\int_{\SL^{2}}\dd g_{1}\dd g_{2}\sum_{\left\{ n_{i}\right\} }\prod_{i=1}^{4}\bra{m_{i}}Y^{\dagger}g_{1}Y\ket{n_{i}}\bra{n_{i}}Y^{\dagger}g_{2}Y\ket{\tilde{m}_{i}}\nonumber \\
\  & \  & =\Lambda^{6(\mu-1)}\T^{2}
\end{eqnarray}
here however the operator $\T$ appear in more general form, since
there are no assumptions on the external spins of the bubble. Thou
detailed study of the operator $\T$ gives an insight to behaviour
of the divergences appearing in the Spin-Foam models.

Finally the the framework of the graph diagrams \cite{ApplicationOSD}
allowed to introduce the simple Feynman-like rules to find the expressions
on the Spin-foam amplitudes (see also \cite{CarloLectures}). These
rules show clearly, that objects $\prod_{i}\bra{m_{i}}Y^{\dagger}gY\ket{m_{i+1}}_{\vec{j}}$
are the main building blocks of amplitude of each diagram (spin-foam).
Thus understanding of them is needed, and the operator $\T=\int_{\SL}\dd g\, Y^{\dagger}gY$
is the first step in this study.

That's why we decided to investigate some basic properties of the
$\T$ operator.

The paper is divided into two parts: the main text and the technical
appendix. In the main text the in \sref{sec:LorentzianPolyhedraPropagator}
there is a strict definition of the object we study. \Sref{sec:Strategy}
contains some remarks on the Saddle Point Approximation method of
integration and points out the issues one has to check to use it in
calculating the Lorentzian Polyhedra Propagator. \Sref{sec:Integrand}
is the study of the function we integrate to obtain the Propagator,
however to make the text easy to read some parts of the calculations
(which are conceptually simple, but technically complicated) are moved
to Appendixes. In \sref{sec:Summary} we gather the results and
point out some possible applications, and in \sref{sec:Conclusions}
we briefly conclude.

There are three appendixes: \ref{sec:AppHypergeometric} is a collection
of useful properties of Gauss Hypergeometric Function $\F$, which
is the main character of the calculations; in \ref{sec:AppSpin} we
present some properties of the $SU(2)$-invariant tensors, and in
\ref{sec:AppProofs} we prove some more technical lemmas.

\section{Lorentzian Polyhedra Propagator\label{sec:LorentzianPolyhedraPropagator}}

To make the paper self-contained, let us fix notation and definitions
and recall some mathematical facts.
\begin{itemize}
\item The $SU(2)$-elements will be denoted usually by $u,v$, the $\SL$-elements
will be denoted by $g$.
\item By $\Hil_{j}$ we will denote the representation space of $SU(2)$
of spin $j$. For the basis elements of $\Hil_{j}$ we will use the
bra-ket notation: $\ket m_{j}$. The basis elements of tensor product
$\Hil_{j_{1}}\otimes\cdots\otimes\Hil_{j_{n}}$ are denoted by $\ket{m_{1},\ldots,m_{n}}_{j_{1}\otimes\cdots\otimes j_{n}}$,
or shortly by $\ket{\vec{m}}_{\vec{j}}$
\item The representation spaces of primary series of $\SL$ we will denote
by $\Hil^{(p,k)}$, and its basis elements in bra-ket notation are
$\ket{j,m}^{(p,k)}$.
\item The Wigner matrices will be denoted by $D^{j}(u)_{m'}^{m}:=\bra mu\ket{m'}_{j}$.
The matrix elements of the primary series of $\SL$ representations
are $D^{(p,k)}(g)_{j',m'}^{j,m}:=\bra{j,m}g\ket{j'm'}^{(p,k)}$.
\item The EPRL map $Y$ is an injection of $\Hil_{j}$ into $\Hil^{(\gamma j,j)}$
given by
\begin{equation}
Y:\Hil_{j}\ni\ket m_{j}\mapsto\ket{j,m}^{(\gamma j,j)}\in\Hil^{(\gamma j,j)}
\end{equation}
(see \cite{CarloLectures,NewLook}\cite{CarloLectures,NewLook}).
It is obvious to generalise the EPRL map to tensor products of $\Hil_{j}$
spaces: $Y:\left(\ket a\otimes\ket b\right)\mapsto Y\ket a\otimes Y\ket b$.
\item %
{} The choice of the basis in $\Hil^{(p,j)}$ picks a normal subgroup
$SU(2)\vartriangleleft\SL$ such that given $u\in SU(2)$ the matrix
elements are $D^{(p,k)}(u)_{j',m'}^{j,m}=\delta_{j'}^{j}D^{j}(u)_{m'}^{m}$
(in other words $u$ commutes with $Y$). This subgroup defines a
decomposition of $\SL$ into $H^{3}\ltimes SU(2)$ (and $SU(2)\rtimes H^{3}$)
where $H^{3}$ is a 3-hyperboloid, which can be parametrised by $\mathbb{R}^{3}$.
\end{itemize}
Having defined above, we can proceed to the Lorentzian Polyhedra Propagator.

\subsection{Definition\label{sub:LPPDefinition}}

Given a set of spinlabels $j_{j},\ldots,j_{N}$ let us define an operator
$\T$ acting on $\Hil_{j_{1}}\otimes\cdots\otimes\Hil_{j_{N}}$ by
the formula
\begin{equation}
\T:=\int_{\SL}\dd g\ Y^{\dagger}gY
\end{equation}

The matrix elements of $\T$ in the spin-$z$ basis can be expressed
as
\begin{equation}
\bra{m_{1},\ldots,m_{N}}\T\ket{m_{1}',\ldots,m_{N}'}_{\vec{j}}=:T_{m_{1}'\cdots m_{N}'}^{m_{1}\cdots m_{N}}=\int_{\SL}\dd g\prod_{i=1}^{N}D^{(\gamma j,j)}(g)_{j_{i}m_{i}'}^{j_{i}m_{i}}
\end{equation}

Let us now study some basic properties of $\T$.

\subsection{Domain and rank\label{sub:LPPDomain}}

Since each $\SL$ element $g$ can be decomposed into $g=k\cdot u$
where $u\in SU(2)$ and $k\in H^{3}$, the $\T$ operator can be written
as
\begin{equation}
\T=\int_{\SL}\dd g\, Y^{\dagger}gY=\int_{H^{3}\ltimes SU(2)}\dd k\,\dd u\, Y^{\dagger}k\cdot u\, Y
\end{equation}
but since the $SU(2)$ elements commute with $Y$:
\begin{equation}
\T=\int_{H^{3}}\dd k\, Y^{\dagger}k\, Y\int_{SU(2)}\dd u\: u
\end{equation}
Now note, that $\int_{SU(2)}\dd u\, u=:P_{{\rm inv}}$ is the projection
onto the $SU(2)$-invariant subspace of the space we are acting on
- in this case it is $\Hil_{{\rm inv}}:={\rm inv}\left(\Hil_{j_{1}}\otimes\cdots\otimes\Hil_{j_{N}}\right)$.
Thus
\begin{equation}
\T=\hat{A}\cdot P_{{\rm inv}}
\end{equation}
 where $\hat{A}$ is some operator.

We can do the same decomposition on the left-hand side:
\begin{eqnarray}
\T & = & \int_{\SL}\dd g\, Y^{\dagger}gY=\int_{SU(2)\rtimes H^{3}}\dd k\,\dd u\, Y^{\dagger}u\cdot k\, Y=\nonumber \\
\  & = & \int_{SU(2)}\dd u\: u\,\int_{H^{3}}\dd k\, Y^{\dagger}k\, Y=P_{{\rm inv}}\cdot\hat{A}
\end{eqnarray}
thus
\begin{equation}
\T=P_{{\rm {\rm inv}}}\cdot\hat{A}\cdot P_{{\rm inv}}
\end{equation}

So $\T$ acts nontrivially only on $\Hil_{{\rm inv}}$. Let us now
choose an orthonormal basis $\left\{ \ket{\iota_{i}}\right\} \in\Hil_{{\rm inv}}$.
Let $\ket{\iota}$ and $\ket{\iota'}$ be the basis elements. It is
enough to investigate the matrix elements of $\T$
\begin{equation}
\T_{\iota'}^{\iota}:=\int_{\SL}\dd g\bra{\iota}Y^{\dagger}g\: Y\ket{\iota'}=\int_{\SL}\dd g\,\Phi_{\iota'}^{\iota}(g)
\end{equation}
what we shell do in what follows.

\subsection{Symmetries\label{sub:LPPSymmetries}}

The vector fields on $\SL$ are spanned by three $SU(2)$ rotation
generators $L^{i}$ and three boost generators $K^{i}$. The $k$
component of the decomposition $g=k\cdot u$ can be written as $k(\vec{\eta})=e^{\vec{\eta}\cdot\vec{K}}$.
For each vector $\vec{\eta}$ (of length $\left|\vec{\eta}\right|=:\eta$)
exists an $SU(2)$ element $u_{\vec{\eta}}$ such that $\frac{\vec{\eta}\cdot\vec{K}}{\eta}=u_{\vec{\eta}}^{\dagger}K^{3}u_{\vec{\eta}}$,
and thou
\begin{equation}
k(\vec{\eta})=e^{\vec{\eta}\cdot\vec{K}}=e^{\eta u_{\vec{\eta}}^{\dagger}K^{3}u_{\vec{\eta}}}=u_{\vec{\eta}}^{\dagger}e^{\eta K^{3}}u_{\vec{\eta}}\label{eq:k_od_eta}
\end{equation}
Thus each $g\in\SL$ can be expressed as 
\begin{equation}
g=u_{\vec{\eta}}^{\dagger}e^{\eta K^{3}}u_{\vec{\eta}}\cdot u
\end{equation}

Let us now investigate, how does $\Phi_{\iota'}^{\iota}(g)$ depend
on $u$ and $\vec{\eta}$.

\subsubsection{$SU(2)$ symmetry}

Using the fact, that $SU(2)$-elements commute with the EPRL map $Y$
it is straightforward to see that
\begin{equation}
\Phi_{\iota'}^{\iota}(k\cdot u)=\bra{\iota}Y^{\dagger}k\cdot u\, Y\ket{\iota'}=\bra{\iota}Y^{\dagger}k\, Y\, u\ket{\iota'}
\end{equation}
but since $\iota'$ is an $SU(2)$-invariant, we have $u\ket{\iota'}=\ket{\iota'}$and
thus
\begin{equation}
\Phi_{\iota'}^{\iota}(k\cdot u)=\bra{\iota}Y^{\dagger}k\, Y\ket{\iota'}=\Phi_{\iota'}^{\iota}(k)
\end{equation}
so the integrand is $SU(2)$ invariant.

\subsubsection{Rotation-of-boost symmetry}

Now using \eref{eq:k_od_eta} lets investigate $\Phi_{\iota'}^{\iota}(k(\vec{\eta}))$:
\begin{equation}
\Phi_{\iota'}^{\iota}(k(\vec{\eta}))=\bra{\iota}Y^{\dagger}u_{\vec{\eta}}^{\dagger}e^{\eta K^{3}}u_{\vec{\eta}}\, Y\ket{\iota'}=\bra{\iota}u_{\vec{\eta}}^{\dagger}Y^{\dagger}e^{\eta K^{3}}Y\, u_{\vec{\eta}}\ket{\iota'}
\end{equation}
again using $SU(2)$ invariance of $\iota$ and $\iota'$ we get
\begin{equation}
\Phi_{\iota'}^{\iota}(k(\vec{\eta}))=\bra{\iota}Y^{\dagger}e^{\eta K^{3}}Y\ket{\iota'}=:\Phi_{\iota'}^{\iota}(\eta)
\end{equation}

Thus the integrand depends only on the length of the boost vector
$\eta$.

\subsection{Integral measure\label{sub:LPPIntegralMeasure}}

Thanks the $SU(2)$-symmetry presented above we can trivially do the
$SU(2)$-integral, (which gives identity, thanks to normalisation
of the Haar measure) and we are left with the integral
\begin{equation}
\T_{\iota'}^{\iota}=\int_{H^{3}}\dd k(\vec{\eta})\,\Phi_{\iota'}^{\iota}(\eta)
\end{equation}
According to \cite{Ruhl} the measure $\dd k$ is
\begin{equation}
\dd k(\vec{\eta})=\frac{1}{\left(4\pi\right)^{2}}\dd\phi(\vec{\eta})\sin\theta(\vec{\eta})\dd\theta(\vec{\eta})\left(\sinh\left|\vec{\eta}\right|\right)^{2}\dd\left|\vec{\eta}\right|
\end{equation}
where $\phi(\vec{\eta})$ and $\theta(\vec{\eta})$ are the spherical
angles of the boost direction.

Let us now introduce the measure function
\begin{equation}
\mu\left(\vec{x}\right):=\left(\frac{\sinh\left|\vec{x}\right|}{4\pi\left|\vec{x}\right|}\right)^{2}
\end{equation}
Now obviously 
\begin{equation}
\dd k\left(\vec{\eta}\right)=\mu\left(\vec{\eta}\right)\dd^{3}\vec{\eta}
\end{equation}
and this is the measure we will use in further calculations, i.e.
we will calculate
\begin{equation}
\T_{\iota'}^{\iota}=\int_{\mathbb{R}^{3}}\dd^{3}\vec{\eta}\,\mu\left(\vec{\eta}\right)\Phi_{\iota'}^{\iota}\left(\left|\vec{\eta}\right|\right)\label{eq:Integral}
\end{equation}
.

\section{Strategy of integration\label{sec:Strategy}}

To calculate the $\T$ operator one have to do the integral over $\SL$
group. We will not do this integral explicitly, we will find its leading
order using the saddle point approximation (SPA) method \cite{SPA}.
The SPA allows to express integrals of the form $\int\dd^{N}x\, g(x)e^{-\Lambda f(x)}$
for large values of $\Lambda$ as a power series in $\frac{1}{\Lambda}$.
The leading order of the integral is determined by the value of integrand
in the critical point $x_{0}$ of the function $f(x)$:
\begin{equation}
\int\dd^{N}x\, g(x)e^{-\Lambda f(x)}=\left(\frac{2\pi}{\Lambda}\right)^{\frac{N}{2}}\left(\left|\frac{\partial^{2}f}{\partial x^{2}}\right|_{x_{0}}\right)^{-\frac{1}{2}}g(x_{0})e^{-\Lambda f(x_{0})}\left(1+O\left(\Lambda^{-1}\right)\right)\label{eq:SPA_definition}
\end{equation}
 where $\left|\frac{\partial^{2}f}{\partial x^{2}}\right|$ is the
determinant of the Hessian matrix of the function $f(x)$. If the
function $f$ has more than one critical point $\{x_{1},\ldots,x_{k}\}$,
than the argument $x_{0}$ that appear in the formula \eref{eq:SPA_definition}
is \emph{the maximal critical point}: such a point $x_{0}\in\{x_{1},\ldots,x_{k}\}$
that $\Re\left(-f(x_{0})\right)$ is maximal (if there is more then
one maximal critical point, than sum over them).

Several assumptions must be satisfied for the formula \eref{eq:SPA_definition}
to be valid. First of all the function $f(x)$ must be smooth and
twice differentiable at the point $x_{0}$. Moreover the integrand
$g(x)e^{-\Lambda f(x)}$ must vanish outside a compact region $\Omega\subset\mathbb{R}^{N}$
(or decay sufficiently fast with $\left|\vec{x}\right|\to\infty$
-- see \sref{sub:SNoncompactness}). These properties
of our integrand $\Phi_{\iota'}^{\iota}(g)$ will be checked in \sref{sec:Integrand}.

\subsection{Non obvious integrand\label{sub:SNonObvious}}

The euclidean EPRL transition amplitude can be easily expressed in
the form of $\int\dd x\, g(x)e^{-\Lambda f(x)}$ \cite{EucleadianEPRL}.
In the Lorentzian case, which we consider now, the decomposition of
the integrand is not obvious. However one can still use the SPA method
for the integrand of the form $\Phi(\Lambda,x)$ defined on $\mathbb{R}_{+}\times\Omega$,
if it has a proper large-$\Lambda$ behaviour and if one can identify
the critical points of the exponent part of the integrand \cite{Kamyk}.

Given a function $\Phi(\Lambda,x)$ let us define the \emph{exponent
part of the integrand:}
\begin{equation}
\phi(x):=\lim_{\Lambda\to\infty}\frac{1}{\Lambda}\log\left(\Phi(\Lambda,x)\right)
\end{equation}
The function $\phi(x)$ may diverge for some points $x$. If the points,
where it diverges, form a region - it means, that the parameter $\Lambda$
is not a good largeness parameter, i.e. the exponent grows faster
then linear in $\Lambda$. In such case one should consider another
parameter $\tilde{\Lambda}(\Lambda)$ instead (this is for example
the case when integrating $\Phi(a,x)=e^{-a^{2}x^{2}}$ - the proper
largeness parameter is $a^{2},$ not $a$). A similar situation is
when the limit is identically zero - then the exponent grows slower
then linearly in $\Lambda$ (for example $\Phi(a,x)=e^{-\sqrt{a}x^{2}}$).
If the largeness parameter $\Lambda$ was chosen properly, we obtain
a nontrivial function $\phi(x)$ (which of course may have some poles
or zero points).

The potentially critical points are the critical points of $\phi$
and the poles of $\phi$. The poles represent the situation when $g(x)=0$.
In this case value of $\phi(x)$ does not capture behaviour of $f$,
hence one need to consider instead
\begin{equation}
\phi^{n}(x):=\lim_{\Lambda\to\infty}\frac{1}{\Lambda}\log\left(\left|\nabla\right|^{n}\Phi(\Lambda,x)\right)
\end{equation}
where $n$ is the lowest order of differentiation, for which the limit
converges, and $\left|\nabla\right|f:=\left|\sum_{i}\partial_{i}f\right|$.
Now let us define the family of sets of poles $B^{(n)}:=\left\{ x\in\Omega\,:\,\phi^{n}(x)=\infty\right\} $
(with $\phi^{0}=\phi$) and a family of sets of critical points $A^{(n)}:=\left\{ x\in B^{(n-1)}\,:\,\nabla\phi^{n}(x)=0\right\} $
(with $B^{(-1)}=\Omega$). The set of critical points of the exponent
is $A:=\bigcup_{n\ge0}A^{(n)}$ .

For each critical point $x_{i}\in A$ let us define the real part
of the exponent as
\begin{equation}
F_{i}:=\Re\left(\phi^{n}(x_{i})\right)\ \ \ {\rm where}\ n\ {\rm such\ that}\ x_{i}\in A^{(n)}
\end{equation}
Now the maximal critical point $x_{\max}$ is the one, for which $F_{i}$
is maximal.

So at the end of the day the integral equals
\begin{equation}
\int_{\Omega}\dd^{N}x\,\Phi(\Lambda,x)=\left(\frac{2\pi}{\Lambda}\right)^{\frac{N}{2}}\left(\left.\frac{\partial^{2}\phi^{n}}{\partial x^{2}}\right|_{x_{\max}}\right)^{-\frac{1}{2}}\Phi(\Lambda,x_{\max})\left(1+O\left(\Lambda^{-1}\right)\right)
\end{equation}
where $n$ is such that $x_{\max}\in A^{(n)}$.

\subsection{Spherically symmetric integrand\label{sub:SSymmetry}}

Consider a spherically symmetric integrand $\Phi\left(\Lambda,\vec{x}\right)=\Phi\left(\Lambda,r\right)$,
for $r=\left|\vec{x}\right|$, with the maximal critical point at
$r=0$ (being in the interior of the region $\Omega$). Then the Hessian
matrix of the function $\phi$ is
\begin{equation}
\frac{\partial^{2}\phi}{\partial x^{i}\partial x^{j}}=\frac{\partial^{2}r}{\partial x^{i}\partial x^{j}}\frac{\dd\phi}{\dd r}+\frac{\partial r}{\partial x^{i}}\frac{\partial r}{\partial x^{j}}\frac{\dd^{2}\phi}{\dd r^{2}}
\end{equation}
Since $r=0$ is the critical point, the differential $\left.\frac{\dd\phi}{\dd r}\right|_{r=0}=0$,
thus the first term vanish.

Further simplification comes from the fact, that 
\begin{equation}
\frac{1}{2}\frac{\partial^{2}r^{2}}{\partial x^{i}\partial x^{j}}=\frac{\partial r}{\partial x^{i}}\frac{\partial r}{\partial x^{j}}+r\frac{\partial^{2}r}{\partial x^{i}\partial x^{j}}=\frac{\partial r}{\partial x^{i}}\frac{\partial r}{\partial x^{j}}\ \ \ {\rm at}\ r=0
\end{equation}
and since $r^{2}=\sum\left(x^{i}\right)^{2}$, we have
\begin{equation}
\frac{\partial r}{\partial x^{i}}\frac{\partial r}{\partial x^{j}}=\delta_{ij}\ \ \ {\rm at}\ r=0
\end{equation}
Thus the Hessian matrix is
\begin{equation}
\left.\frac{\partial^{2}\phi}{\partial x^{i}\partial x^{j}}\right|_{\vec{x}=0}=\left.\frac{\dd^{2}\phi}{\dd r^{2}}\right|_{r=0}\delta_{ij}
\end{equation}
And the Hessian determinant
\begin{equation}
\left|\frac{\partial^{2}\phi}{\partial x^{i}\partial x^{j}}\right|_{\vec{x}=0}=\left|\frac{\dd^{2}\phi}{\dd r^{2}}\right|_{r=0}^{N}
\end{equation}

\subsection{Non compactness of integration range\label{sub:SNoncompactness}}

The SPA method is well defined if the region of integration $\Omega$
is compact and the integrand vanish at $\partial\Omega$. However
under certain assumptions one can generalise the SPA to the case of
noncompact integration range. Assume that exists $\Lambda_{0}$ such
that for all $\Lambda>\Lambda_{0}$ the following is true: for each
$\epsilon>0$ exists a compact region $R_{\epsilon}\subset\Omega$,
such that
\begin{equation}
\int_{\Omega\setminus R_{\epsilon}}\dd^{N}x\,\left|\Phi(\Lambda,x)\right|<\epsilon
\end{equation}
and that $R_{\epsilon}\subset R_{\epsilon'}$ for $\epsilon>\epsilon'$.
Then let us introduce for each $\epsilon$ another compact region
$\tilde{R}_{\epsilon}$, such that $R_{\epsilon}\subsetneq\tilde{R}_{\epsilon}\subset\Omega$
and a smooth function $\chi_{\epsilon}(x)$ such that
\begin{equation}
\chi_{\epsilon}(x)=\begin{cases}
1 & x\in R_{\epsilon}\\
0\le\chi_{\epsilon}(x)\le1 & x\in\tilde{R}_{\epsilon}\setminus R_{\epsilon}\\
0 & x\in\Omega\setminus\tilde{R}_{\epsilon}
\end{cases}
\end{equation}
Obviously $\int_{\Omega\setminus R_{\epsilon}}\dd^{N}x\,\left|\chi_{\epsilon}(x)\Phi(\Lambda,x)\right|<\epsilon$.
Now let $I(\Lambda):=\int_{\Omega}\dd^{N}x\Phi(\Lambda,x)$ and $I_{\epsilon}(\Lambda):=\int_{\Omega}\dd^{N}x\,\chi_{\epsilon}(x)\Phi(\Lambda,x)$.
Obviously for all $\Lambda>\Lambda_{0}$ we have $\left|I(\Lambda)-I_{\epsilon}(\Lambda)\right|<2\epsilon$,
thus function $I_{\epsilon}(\Lambda)$ converges to $I(\Lambda)$
uniformly with respect to $\Lambda$. But each integral $I_{\epsilon}(\Lambda)$
is in fact integral over a compact region $\tilde{R}_{\epsilon}$,
so it can be calculated using the SPA method:
\begin{equation}
I_{\epsilon}(\Lambda)=\left(\frac{2\pi}{\Lambda}\right)^{\frac{N}{2}}\left(\left|\frac{\partial^{2}\phi}{\partial x^{2}}\right|_{x_{\epsilon}}\right)^{-\frac{1}{2}}\Phi(\Lambda,x_{\epsilon})\left(1+O\left(\Lambda^{-1}\right)\right)
\end{equation}
where $x_{\epsilon}$ is the maximal critical point of $\phi$ in
the region $R_{\epsilon}$. For $\epsilon$ sufficiently small, for
example for $\epsilon<\frac{1}{\Lambda}\left(\frac{2\pi}{\Lambda}\right)^{\frac{N}{2}}\left(\left|\frac{\partial^{2}\phi}{\partial x^{2}}\right|_{x_{0}}\right)^{-\frac{1}{2}}\Phi(\Lambda,x_{0})$,
the region $R_{\epsilon}$ must contain the maximal critical point
of $\phi$, so $x_{\epsilon}=x_{0}$, thus the leading term does not
depend on $\epsilon$. So the leading term of the function $I(\Lambda)$
is the limit at $\epsilon\to0$ of the leading terms of $I_{\epsilon}(\Lambda)$.

\subsection{Multiplication by a $\Lambda$-independent function\label{sub:SMeasure}}

Consider now integral of the form 
\begin{equation}
\tilde{I}(\Lambda):=\int_{\Omega}\dd^{N}x\,\mu(\vec{x})\Phi(\Lambda,\vec{x})
\end{equation}
where $\mu(\vec{x})$ is a nonvanishing function. When using the SPA
method, the integral $\tilde{I}(\Lambda)$ can be easily related to
$I(\Lambda):=\int_{\Omega}\dd^{N}x\,\Phi(\Lambda,x)$. Indeed, note,
that presence of $\mu(\vec{x})$ does not effect the exponent part
of the integrand:
\begin{equation}
\tilde{\phi}(\vec{x})=\lim_{\Lambda\to\infty}\frac{1}{\Lambda}\ln\mu(\vec{x})+\frac{1}{\Lambda}\ln\Phi(\Lambda,\vec{x})=0+\lim_{\Lambda\to\infty}\frac{1}{\Lambda}\ln\Phi(\Lambda,\vec{x})=\phi(\vec{x})
\end{equation}
thus the maximal critical points of the exponent of the integrand
does not know about presence of $\mu(\vec{x})$. So using SPA method
we get
\begin{equation}
\tilde{I}(\Lambda)=\left(\frac{2\pi}{\Lambda}\right)^{\frac{N}{2}}\left(\left.\frac{\partial^{2}\phi^{n}}{\partial x^{2}}\right|_{x_{\max}}\right)^{-\frac{1}{2}}\mu(x_{\max})\Phi(\Lambda,x_{\max})\left(1+O\left(\Lambda^{-1}\right)\right)=\mu(x_{\max})I(\Lambda)
\end{equation}

\section{Detailed analysis of the integrand\label{sec:Integrand}}

\subsection{Measure and exponent part of the integrand}

Let us now investigate the properties of the integrand. Recalling
\eref{eq:Integral} the integrand is $\tilde{\Phi}(\eta)=\mu(\eta)\Phi_{\iota'}^{\iota}(\eta)$
with
\begin{equation}
\mu(\eta)=\left(\frac{\sinh\eta}{4\pi\eta}\right)^{2}\ \ \ \ \Phi_{\iota'}^{\iota}(\eta)=\bra{\iota}Y^{\dagger}e^{\eta K^{3}}Y\ket{\iota'}
\end{equation}
and $\iota,\iota'\in{\rm inv}\left(\Hil_{j_{1}}\otimes\cdots\otimes\Hil_{j_{N}}\right)$.
Our largeness parameter is $J=\max\left\{ j_{i}\right\} _{i=1,\ldots,N}$
. Obviously the function $\mu(\eta)$ does not depend on the largeness
parameter. Thus, using \sref{sub:SMeasure}, it is enough
to find the critical points of $\Phi_{\iota'}^{\iota}(\eta)$. From
now on we will call the function $\tilde{\Phi}(\eta)$ the \emph{full
integrand}, the function $\Phi_{\iota'}^{\iota}(\eta)$ the \emph{integrand},
and the function $\mu(\eta)$ the \emph{measure part of the integrand}
or simply the \emph{measure}.

In this section we will often go back to the $\ket m_{j}$ basis,
so that
\begin{equation}
\Phi_{\iota'}^{\iota}(\eta)=\sum_{\vec{m}\vec{m'}}\bra{\iota}\ket{\vec{m}}_{\vec{j}}\bra{\vec{m'}}\ket{\iota'}_{\vec{j}}\Phi_{\vec{m'}}^{\vec{m}}(\eta)
\end{equation}
 where
\begin{equation}
\Phi_{\vec{m'}}^{\vec{m}}(\eta)=\prod_{i=1}^{N}\bra{m_{i}}Y^{\dagger}e^{\eta K^{3}}Y\ket{m_{i}'}_{j_{i}}
\end{equation}
Since $\left[K^{3},L^{3}\right]=0$, each term is proportional to
$\delta_{m_{i},m_{i}'}$, and thus we can define the function 
\begin{equation}
f_{m}^{(j)}(\eta):=\bra mY^{\dagger}e^{\eta K^{3}}Y\ket m_{j}
\end{equation}
 such that
\begin{equation}
\Phi_{\vec{m'}}^{\vec{m}}(\eta)=\prod_{i=1}^{N}\delta_{m_{i},m_{i}'}f_{m_{i}}^{(j_{i})}(\eta)\label{eq:PhiJako_f}
\end{equation}
Thanks to \eref{eq:PhiJako_f} it is now obvious, that in the basis
$\ket{\vec{m}}_{\vec{j}}$ all the nondiagonal matrix elements are
precisely zero (without any approximation). 

The exponent part of the integrand is
\begin{equation}
\phi_{\iota'}^{\iota}(\eta)=\lim_{J\to\infty}\phi_{\iota'}^{\iota}(\eta,J):=\lim_{J\to\infty}\frac{1}{J}\ln\left[\sum_{\vec{m}}\overline{\iota_{\vec{m}}}\iota'_{\vec{m}}e^{\sum_{i=1}^{N}\ln f_{m_{i}}^{(j_{i})}(\eta)}\right]
\end{equation}
Because of the sum under the logarithm it is quite inconvenient object,
thus most calculations we will do for the exponent part of the function
$\Phi_{\vec{m'}}^{\vec{m}}(\eta)$, namely:

\begin{equation}
\phi_{\vec{m}}(\eta)=\lim_{J\to\infty}\phi_{\vec{m}}(\eta,J):=\lim_{J\to\infty}\frac{1}{J}\sum_{i=1}^{N}\ln f_{m_{i}}^{(j_{i})}(\eta)
\end{equation}
Note, that
\begin{equation}
\phi_{\iota'}^{\iota}(\eta,J)=\frac{1}{J}\ln\left[\sum_{\vec{m}}\overline{\iota_{\vec{m}}}\iota'_{\vec{m}}e^{J\phi_{\vec{m}}(\eta,J)}\right]\label{eq:phi_iota_as_phi_m}
\end{equation}

For later convenience let us introduce parameters $x_{i}:=\frac{j_{i}}{J}$.
Obviously $0\le x_{i}\le1$, but one cannot treat them as the small
parameters.

\subsection{Hypergeometric representation\label{sub:IHypergeometric}}

Let us now focus on the function $f_{m}^{(j)}(\eta)$. According to
\cite{Ruhl} it can be written as
\begin{eqnarray}
f_{m}^{(j)}(\eta) & = & \left(2j+1\right)\binom{2j}{j+m}\ e^{-\left(j+m+1\right)\eta}e^{\im\gamma j\eta}\times\label{eq:Definicja_fjm}\\
\  & \  & \times\int_{0}^{1}\dd t\, t^{j+m}\left(1-t\right)^{j-m}\left[1-\left(1-e^{-2\eta}\right)t\right]^{\im\gamma j-\left(j+1\right)}\nonumber 
\end{eqnarray}

Recall now the integral definition of the Gauss's hypergeometric function:
\begin{equation}
\FF abcz=\frac{\Gamma(c)}{\Gamma(b)\Gamma(c-b)}\int_{0}^{1}\dd t\, t^{b-1}\left(1-t\right)^{c-b-1}\left(1-zt\right)^{-a}\label{eq:IntegralDefinition2F1}
\end{equation}
Comparison of formulas \eref{eq:Definicja_fjm} and \eref{eq:IntegralDefinition2F1}
gives a conclusion, that
\begin{equation}
f_{m}^{(j)}(\eta)=e^{-(j+m+1)\eta}e^{\im j\gamma\eta}\FF{j+1-\im\gamma j}{j+1+m}{2j+2}{1-e^{-2\eta}}
\end{equation}

\subsection{Maximum point\label{sub:IMaximumPoint}}

To apply the SPA method we need to find the maximal critical point
of $\phi_{\iota'}^{\iota}(\eta)$. The natural candidate is $\eta=0$.
Since $\left.f_{m}^{(j)}(\eta)\right|_{\eta=0}=1$ (see \eref{eq:Series2F1}),
it is obvious to find the value of the integrand
\begin{equation}
\Phi_{\vec{m}}^{\vec{m}}(0)=1\ \ \ {\rm so}\ \ \ \Phi_{\iota'}^{\iota}=\bra{\iota}\ket{\iota'}
\end{equation}

Now let us show, that $\eta=0$ is the only critical point that counts,
i.e. was there any other critical point $\eta_{1}$, The value of
$\Phi_{\vec{m}}^{\vec{m}}(\eta_{1})$ would be exponentially suppressed
in large $J$ limit. Indeed, one can estimate the modulus of $f_{m}^{(j)}(\eta)$
by
\begin{equation}
\left|f_{m}^{(j)}(\eta)\right|\le\left(e^{1-2\eta-e^{-2\eta}}\right)^{\frac{(j+1)^{2}-m^{2}}{4(2j+3)}}
\end{equation}
(see \ref{sub:AHEstymations}) thus
\begin{equation}
\left|\Phi_{\vec{m}}^{\vec{m}}(\eta_{1})\right|\le\left(e^{1-2\eta_{1}-e^{-2\eta_{1}}}\right)^{\sum_{i=1}^{N}\frac{(j_{i}+1)^{2}-m_{i}^{2}}{4(2j_{i}+3)}}=\left[C(\eta_{1})\right]^{\sum_{i=1}^{N}\frac{(j_{i}+1)^{2}-m_{i}^{2}}{4(2j_{i}+3)}}
\end{equation}
Now since
\begin{equation}
\left|\Phi_{\iota'}^{\iota}(\eta_{1})\right|=\left|\sum_{\vec{m}}\Phi_{\vec{m}}^{\vec{m}}(\eta_{1})\overline{\iota_{\vec{m}}}\iota'_{\vec{m}}\right|\le\left|\sum_{\vec{m}}\left|\Phi_{\vec{m}}^{\vec{m}}(\eta_{1})\right|\overline{\iota_{\vec{m}}}\iota'_{\vec{m}}\right|
\end{equation}
we have
\begin{equation}
\left|\Phi_{\iota'}^{\iota}(\eta_{1})\right|\le\left|\sum_{\vec{m}}\left[C(\eta_{1})\right]^{\sum_{i=1}^{N}\frac{(j_{i}+1)^{2}-m_{i}^{2}}{4(2j_{i}+3)}}\overline{\iota_{\vec{m}}}\iota'_{\vec{m}}\right|
\end{equation}
Now using the fact, that the states $\ket{\iota}$ and $\ket{\iota'}$
are $SU(2)$-gauge invariant we can replace $\sum_{i=1}^{N}m_{i}^{2}$
by $\sum_{i=1}^{N}\frac{j_{i}\left(j_{i}+1\right)}{3}$ in the formula
above (see \ref{sec:AppSpin}) and obtain 
\begin{equation}
\left|\Phi_{\iota'}^{\iota}(\eta_{1})\right|\le\left|\bra{\iota}\ket{\iota'}\right|\left[C(\eta_{1})\right]^{\frac{1}{12}\sum_{i=1}^{N}j_{i}+1}=\left|\bra{\iota}\ket{\iota'}\right|\left[C(\eta_{1})\right]^{\frac{1}{12}J\sum_{i=1}^{N}x_{i}+\frac{1}{J}}\label{eq:PhiEstimation}
\end{equation}
(see \eref{eq:AppEstimationOfPhi}). Since $\forall_{\eta_{1}>0}C(\eta_{1})<1$
(see\eref{sub:APInequality}), it is obvious, that 
\begin{equation}
\left|\Phi_{\iota'}^{\iota}(\eta_{1})\right|\ll\left|\Phi_{\iota'}^{\iota}(0)\right|\ \ \ {\rm for}\ J\gg1
\end{equation}

\subsection{Smoothness check\label{sub:ISmoothnessCheck}}

Having proven that the maximal critical point of $\phi_{\iota'}^{\iota}$
is at $\eta=0$ (if any), let us check, whether the exponent part
of the integrand is smooth at this point, i.e. if $\left.\frac{\dd\phi_{\iota'}^{\iota}}{\dd\eta}\right|_{\eta=0}=0$?
Using the property \eref{eq:phi_iota_as_phi_m} we have
\begin{equation}
\left.\frac{\dd\phi_{\iota'}^{\iota}}{\dd\eta}\right|_{\eta=0}=\left.\frac{\frac{1}{J}\sum_{\vec{m}}\overline{\iota_{\vec{m}}}\iota'_{\vec{m}}Je^{J\phi_{\vec{m}}(\eta)}\frac{\dd\phi_{\vec{m}}}{\dd\eta}}{\Phi_{\iota'}^{\iota}(\eta)}\right|_{\eta=0}=\sum_{\vec{m}}\frac{\overline{\iota_{\vec{m}}}\iota'_{\vec{m}}}{\bra{\iota}\ket{\iota'}}\left.\frac{\dd\phi_{\vec{m}}}{\dd\eta}\right|_{\eta=0}\label{eq:FirstDerivative_phi_iota}
\end{equation}
We will show, that $\left.\frac{\dd\phi_{\vec{m}}}{\dd\eta}\right|_{\eta=0}=0$
for all $\vec{m}$.

Let us then calculate:
\begin{equation}
\frac{\partial}{\partial\eta}\phi_{\vec{m}}(\eta,J)=\frac{1}{J}\sum_{i=1}^{N}\frac{\frac{\dd f_{m_{i}}^{(j_{i})}}{\dd\eta}}{f_{m_{i}}^{(j_{i})}(\eta)}
\end{equation}
Details of calculations can be found in \ref{sub:AHBasicFirst}, the
result is
\begin{equation}
\frac{\partial}{\partial\eta}\phi_{\vec{m}}(\eta,J)=\frac{1}{J}\left[\sum_{i=1}^{N}-\delta_{i}+2\frac{a_{i}b_{i}}{c_{i}}e^{-2\eta}\frac{\FF{a_{i}+1}{b_{i}+1}{c_{i}+1}{1-e^{-2\eta}}}{\FF{a_{i}}{b_{i}}{c_{i}}{1-e^{-2\eta}}}\right]
\end{equation}
with $\delta_{i}=\delta:=j_{i}+1+m_{i}-\im\gamma j_{i}$, $a_{i}=1+j_{i}-\im\gamma j_{i}$,
$b_{i}=1+j_{i}+m_{i}$, $c_{i}=2j_{i}+2$. For $\eta=0$ all the hypergeometric
functions have value $1$, and we are left with
\begin{equation}
\left.\frac{\partial}{\partial\eta}\phi_{\vec{m}}(\eta,J)\right|_{\eta=0}=-\frac{1}{J}\sum_{i=1}^{N}\delta_{i}-2\frac{a_{i}b_{i}}{c_{i}}
\end{equation}
using simple algebra (similarly to \eref{eq:dfjmn0}) we can simplify
it to:
\begin{equation}
\left.\frac{\partial}{\partial\eta}\phi_{\vec{m}}(\eta,J)\right|_{\eta=0}=-\im\gamma\frac{\sum_{i=1}^{N}m_{i}}{J}+O\left(J^{-1}\right)
\end{equation}
Thanks to $SU(2)$ invariance of the states we act on, the sum $\sum_{i=0}^{N}m_{i}=0$.
The second term vanish when the limit $J\to\infty$ is taken. Thus
for all $\vec{m}$
\begin{equation}
\left.\frac{\dd}{\dd\eta}\phi_{\vec{m}}(\eta)\right|_{\eta=0}=0
\end{equation}
so 
\begin{equation}
\left.\frac{\dd\phi_{\iota'}^{\iota}}{\dd\eta}\right|_{\eta=0}=0
\end{equation}
so the exponent part of the integrand is smooth at the maximal critical
point.

\subsection{Asymptotics check\label{sub:IAsymptoticsCheck}}

As we have already shown in \ref{eq:PhiEstimation}, the modulus of
the integrand is bounded by
\begin{equation}
\left|\Phi_{\iota'}^{\iota}(\eta)\right|\le\left(e^{1-2\eta-e^{-2\eta}}\right)^{\frac{J}{12}\sum_{i=1}^{N}x_{i}+\frac{1}{J}}
\end{equation}
Thus and for arbitrary small $\epsilon$ one can find such an $\eta_{\epsilon}$
that
\begin{equation}
\forall_{J>1}\ \ \ \left|\int_{\eta_{\epsilon}}^{\infty}\dd\eta\,\mu(\eta)\Phi_{\iota'}^{\iota}(\eta)\right|<\epsilon
\end{equation}
Indeed, this integral can be estimated by
\begin{eqnarray}
\left|\int_{\eta_{\epsilon}}^{\infty}\dd\eta\,\mu(\eta)\Phi_{\iota'}^{\iota}(\eta)\right| & \le & \int_{\eta_{\epsilon}}^{\infty}\dd\eta\,\left|\mu(\eta)\Phi_{\iota'}^{\iota}(\eta)\right|\nonumber \\
\  & \le & \int_{\eta_{\epsilon}}^{\infty}\dd\eta\,\left(\frac{\sinh\eta}{4\pi\eta}\right)^{2}\left(e^{1-2\eta-e^{-2\eta}}\right)^{\frac{J}{12}\sum_{i=1}^{N}\left(x_{i}+\frac{1}{J}\right)}\label{eq:phi_estimation}\\
\  & \le & \frac{1}{2}\left(\frac{e}{4\pi\eta_{\epsilon}}\right)^{2}\left(e-e^{1-e^{-2\eta_{\epsilon}}}-e^{1-2\eta_{\epsilon}-e^{-2\eta_{\epsilon}}}\right)\nonumber 
\end{eqnarray}
where the last inequality holds for $\frac{J}{12}\sum_{i=1}^{N}\left(x_{i}+\frac{1}{J}\right)$>3
(see \ref{sub:APEstimation}) - what is for sure true for $J>36$,
and since we consider the large $J$ limit, this condition holds in
the limit. As it is shown in \ref{sub:APExistence}, the inequality
$\frac{1}{2}\left(\frac{e}{4\pi\eta_{\epsilon}}\right)^{2}\left(e-e^{1-e^{-2\eta_{\epsilon}}}-e^{1-2\eta_{\epsilon}-e^{-2\eta_{\epsilon}}}\right)<\epsilon$
has always a solution. Thus the assumptions of the lemma of
\sref{sub:SNoncompactness} are satisfied, so our integrand $\Phi_{\iota'}^{\iota}$
has proper asymptotic behaviour and the SPA method can be applied.

\subsection{Second derivative\label{sub:ISecondDerivative}}

We need to know the Hessian determinant $\left|\frac{\partial^{2}\phi_{\iota'}^{\iota}}{\partial\vec{\eta}^{2}}\right|$
at $\eta=0$, which thanks to the spherical symmetry is equal $\left|\frac{\dd^{2}\phi_{\iota'}^{\iota}}{\dd\eta^{2}}\right|_{\eta=0}^{3}$
(see \ref{sub:SSymmetry}).

To calculate the second derivative of $\phi_{\iota'}^{\iota}(\eta)$
we will use the equation \eref{eq:FirstDerivative_phi_iota}:
\begin{eqnarray}
\frac{\dd^{2}\phi_{\iota'}^{\iota}}{\dd\eta^{2}} & = & \frac{\dd}{\dd\eta}\frac{\sum_{\vec{m}}\overline{\iota_{\vec{m}}}\iota'_{\vec{m}}e^{J\phi_{\vec{m}}(\eta)}\frac{\dd\phi_{\vec{m}}}{\dd\eta}}{\Phi_{\iota'}^{\iota}(\eta)}\nonumber \\
\  & = & \sum_{\vec{m}}\frac{\overline{\iota_{\vec{m}}}\iota'_{\vec{m}}e^{J\phi_{\vec{m}}(\eta)}}{\Phi_{\iota'}^{\iota}(\eta)}\frac{\dd^{2}\phi_{\vec{m}}}{\dd\eta^{2}}-\frac{\dd\Phi_{\iota'}^{\iota}}{\dd\eta}\sum_{\vec{m}}\frac{\overline{\iota_{\vec{m}}}\iota'_{\vec{m}}e^{J\phi_{\vec{m}}(\eta)}}{\left[\mbox{\ensuremath{\Phi_{\iota'}^{\iota}}(\ensuremath{\eta})}\right]^{2}}\frac{\dd\phi_{\vec{m}}}{\dd\eta}\label{eq:SecondDerivative_phi}
\end{eqnarray}
The second term in above formula vanish at $\eta=0$. Indeed, in \sref{sub:ISmoothnessCheck}
we have checked, that $\forall_{\vec{m}}\detazero{\phi_{\vec{m}}}=0$
up to $O\left(J^{-1}\right)$ corrections. Thus noting, that $\Phi_{\iota'}^{\iota}(0)=\bra{\iota}\ket{\iota'}$
and $\phi_{\vec{m}}(0)=0$, we have
\begin{equation}
\detadwazero{\phi_{\iota'}^{\iota}}=\frac{1}{\bra{\iota}\ket{\iota'}}\sum_{\vec{m}}\overline{\iota_{\vec{m}}}\iota'_{\vec{m}}\detadwazero{\phi_{\vec{m}}}\label{eq:DetaDwa_phi_zero}
\end{equation}

Let us then analyse $\detadwazero{\phi_{\vec{m}}}$. Obviously it
decomposes into a sum
\begin{equation}
\detadwazero{\phi_{\vec{m}}}=\sum_{i=1}^{N}\frac{1}{J}\detadwazero{\ln\left[\fjmi\right]}
\end{equation}

Using calculations of \ref{sub:AHBasicSecond} we get (up to $\frac{1}{J}$
corrections)
\begin{equation}
\detadwazero{\ln\left[\fjmi\right]}=-\left(1+\gamma^{2}\right)\frac{\left(j_{i}+1\right)^{2}-m_{i}^{2}}{\left(2j_{i}+3\right)}\left(1+O\left(J^{-1}\right)\right)
\end{equation}
so (neglecting the $O\left(J^{-1}\right)$ terms)
\begin{equation}
\detadwazero{\phi_{\vec{m}}}=-\frac{1+\gamma^{2}}{J}\sum_{i=1}^{N}\frac{\left(j_{i}+1\right)^{2}-m_{i}^{2}}{\left(2j_{i}+3\right)}
\end{equation}

Now recalling \eref{eq:DetaDwa_phi_zero} we have
\begin{equation}
\detadwazero{\phi_{\iota'}^{\iota}}=\frac{1}{\bra{\iota}\ket{\iota'}}\sum_{\vec{m}}\left[\overline{\iota_{\vec{m}}}\iota'_{\vec{m}}\left(-\frac{1+\gamma^{2}}{J}\right)\sum_{i=1}^{N}\frac{\left(j_{i}+1\right)^{2}-m_{i}^{2}}{\left(2j_{i}+3\right)}\right]
\end{equation}
Thanks to $SU(2)$ symmetry we may use the lemma of \ref{sec:AppSpin}
and substitute each term $m_{i}^{2}$ by $\frac{j_{i}(j_{i}+1)}{3}$
(see \eref{eq:AppDetaDwa_phi_zero}) and then we get
\begin{equation}
\detadwazero{\phi_{\iota'}^{\iota}}=-\frac{1+\gamma^{2}}{J}\frac{\sum_{\vec{m}}\overline{\iota_{\vec{m}}}\iota'_{\vec{m}}}{\bra{\iota}\ket{\iota'}}\sum_{i=1}^{N}\frac{\left(j_{i}+1\right)}{3}
\end{equation}
 The $\iota$-dependent factor cancels out, because $\sum_{\vec{m}}\overline{\iota_{\vec{m}}}\iota'_{\vec{m}}=\bra{\iota}\ket{\iota'}$:
\begin{equation}
\detadwazero{\phi_{\iota'}^{\iota}}=-\frac{1+\gamma^{2}}{3}\left(\sum_{i=1}^{N}x_{i}+\frac{1}{J}\right)
\end{equation}
Neglecting the $\frac{1}{J}$-order terms simplify the formula a lot
and gives
\begin{equation}
\detadwazero{\phi_{\iota'}^{\iota}}=-\frac{1+\gamma^{2}}{3}\sum_{i=1}^{N}x_{i}
\end{equation}
So the Hessian determinant is
\begin{equation}
\left|\frac{\partial^{2}\phi_{\iota'}^{\iota}}{\partial\vec{\eta}^{2}}\right|_{\eta=0}=\left[\frac{1+\gamma^{2}}{3}\sum_{i=1}^{N}x_{i}\right]^{3}\label{eq:Hesian}
\end{equation}

\section{Summary and applications\label{sec:Summary}}

\subsection{Lorentzian Polyhedra Propagator\label{sub:SumLPP}}

Let us summarise all the calculations done in sections \ref{sec:Strategy}
and \ref{sec:Integrand}. The Lorentzian Polyhedra Propagator $\T$
given by the integral
\begin{equation}
\T_{\iota'}^{\iota}=\int_{\mathbb{R}^{3}}\dd^{3}\vec{\eta}\,\mu(\eta)\Phi_{\iota'}^{\iota}(\eta)
\end{equation}
for $J\gg1$ can by approximated by the value of integrand at $\eta=0$
\begin{equation}
\T_{\iota'}^{\iota}=\left(\frac{2\pi}{J}\right)^{\frac{3}{2}}\left|\frac{\partial^{2}\phi}{\partial\vec{\eta}^{2}}\right|_{\eta=0}^{-\frac{1}{2}}\mu(0)\Phi_{\iota'}^{\iota}(0)\left(1+O\left(J^{-1}\right)\right)
\end{equation}
Noting that $\mu(0)=\left(\frac{1}{4\pi}\right)^{2}$, $\Phi_{\iota'}^{\iota}(0)=\sum_{\vec{m},\vec{m'}}\overline{\iota_{\vec{m}}}\iota_{\vec{m'}}\delta_{\vec{m},\vec{m'}}=\bra{\iota}\ket{\iota'}$,
taking the value of $\detadwazero{\phi}$ from \eref{eq:Hesian}
and neglecting the $\frac{1}{J}$ corrections we get
\begin{equation}
\T_{\iota'}^{\iota}=\left(\frac{1}{4\pi}\right)^{2}\left(\frac{2\pi}{J}\right)^{\frac{3}{2}}\left[\frac{3}{\left(1+\gamma^{2}\right)\sum_{i=1}^{N}x_{i}}\right]^{\frac{3}{2}}\bra{\iota}\ket{\iota'}
\end{equation}
This means that the leading order of $\T$ is proportional to identity:
\begin{equation}
\T=\left(\frac{1}{4\pi}\right)^{2}\left(\frac{2\pi}{J}\right)^{\frac{3}{2}}\left[\frac{3}{\left(1+\gamma^{2}\right)\sum_{i=1}^{N}x_{i}}\right]^{\frac{3}{2}}\mathbb{I}_{\vec{j}}
\end{equation}

The higher order terms of $\T$ are not known yet, but we have already
found (see \eref{eq:PhiJako_f}), that in the basis $\ket{\vec{m}}_{\vec{j}}$
only the diagonal terms are nonzero:
\begin{equation}
\T_{\vec{m'}}^{\vec{m}}=\left(\frac{1}{4\pi}\right)^{2}\left(\left[\frac{6\pi}{J\left(1+\gamma^{2}\right)\sum_{i=1}^{N}x_{i}}\right]^{\frac{3}{2}}+\left(\frac{1}{J}\right)^{\frac{5}{2}}T_{\vec{m}}\right)\delta_{\vec{m'}}^{\vec{m}}
\end{equation}

\subsection{Application in Dipole Cosmology\label{sub:SumDC}}

According to \cite{DC_Lorentzian} the DC transition amplitude is
\begin{eqnarray}
W(z) & = & \sum_{\left\{ j_{\ell}\right\} }\prod_{\ell=1}^{4}\left(2j_{\ell}+1\right)e^{-2t\hbar j_{\ell}(j_{\ell}+1)-\im\lambda v_{0}j_{\ell}^{\frac{3}{2}}-\im zj_{\ell}}\int_{\SL}\dd g\prod_{\ell=1}^{4}\bra{j_{\ell}}u_{\vec{n_{\ell}}}^{\dagger}Y^{\dagger}g\, Y\, u_{\vec{n'_{\ell}}}\ket{j_{\ell}}_{j_{\ell}}\nonumber \\
\  & = & \sum_{\left\{ j_{\ell}\right\} }\prod_{\ell=1}^{4}\left(2j_{\ell}+1\right)e^{-2t\hbar j_{\ell}(j_{\ell}+1)-\im\lambda v_{0}j_{\ell}^{\frac{3}{2}}-\im zj_{\ell}}\,\bra{\iota}\T\ket{\iota'}\\
\  & \  & \ \ \ {\rm with}\ \ \iota^{(')}:=\int_{SU(2)}\dd u\,\prod_{\ell=1}^{4}u\cdot u_{\vec{n^{(')}}}\ket{j_{\ell}}\nonumber 
\end{eqnarray}
Our calculations influence only the term $\bra{\iota}\T\ket{\iota'}$.
Authors assume that it behaves like $\frac{N_{0}}{j_{0}^{3}}$ (with
$j_{0}=\frac{\Im z}{4t\hbar}$ and $N_{0}$ being a constant for a
given graph). The direct formula for the Lorentzian Polyhedra propagator
shows, that it is 
\begin{equation}
\bra{\iota}\T\ket{\iota'}=\tilde{N}\left(x_{i}\right)\cdot J^{-\frac{3}{2}}\bra{\iota}\ket{\iota'}
\end{equation}
where the factor $\tilde{N}\left(x_{i}\right)$ depend only on ratios
of spins $x_{i}=\frac{j_{i}}{J}$, but it does not scale with $J$.
The coherent $SU(2)$-intertwiner have norm squared scaling as $j^{-\frac{3}{2}}$
(see \cite{L-S}), thus the overall scaling of $\bra{\iota}\T\ket{\iota'}$
is $N\cdot j^{-3}$, as expected. So taking into account the direct
calculation of the Lorentzian Polyhedra Propagator, the Dipole Cosmology
transition amplitude reproduce the original DC formula of \cite{DC1,DC_Lorentzian}:
\begin{equation}
W(z)=\sum_{\left\{ j_{\ell}\right\} }\frac{N\left(x_{i}\right)}{j_{0}^{3}}\prod_{\ell=1}^{4}\left(2j_{\ell}+1\right)e^{-2t\hbar j_{\ell}(j_{\ell}+1)-\im\lambda v_{0}j_{\ell}^{\frac{3}{2}}-\im zj_{\ell}}
\end{equation}
with a minor correction: the graph-shape dependent factor $N$ now
depends also on ratios of spins.

Using the same summation technique, as in \cite{DC1}, (i.e. summing
by integrating the Gaussian integrals) one obtains the same result
\begin{equation}
W(z)=\frac{N_{0}}{j_{0}^{3}}\left(2j_{0}\sqrt{\frac{\pi}{t}}e^{-\frac{z^{2}}{8t\hbar}}\right)^{4}=\mathcal{N}ze^{-\frac{z^{2}}{2t\hbar}}
\end{equation}
with $N_{0}$ being the value of $N\left(x_{i}\right)$ factor calculated
for all $x_{i}=1$.

\subsection{Application in radiative corrections\label{sub:SumRadiative}}

As we have already recalled, the radiative corrections to the spin-foam
edge are proportional to the $\T^{2}$ operator\cite{Aldo}. To be
precise, given a cutoff $\Lambda$ on the internal spins of the bubble,
the leading order of the transition amplitude are the matrix elements
of the following operator 
\begin{equation}
W^{\Lambda}=\Lambda^{6(\mu-1)}\T^{2}
\end{equation}
where $\mu$ is the parameter dependent on the model chosen (i.e.
it is a degree of the polynomial that is used to define the face amplitude).
For $\mu=1$ the exponent is zero, which means, that the divergence
is $O\left(\ln(\Lambda)\right)$. Our calculation shows, that in large
$J$ limit (with $J$ being the maximal external spin of the external
faces of the bubble) up to $\frac{1}{J}$ corrections this radiative-corrected
edge operator is proportional to the identity with the constant dependent
on $\Lambda$ and $J$:
\begin{equation}
W_{\iota,\iota'}^{\Lambda}=\ln\Lambda\cdot\left(\frac{1}{J}\right)^{3}\cdot\frac{27}{32\pi}\left[\frac{1}{\left(1+\gamma^{2}\right)\sum_{i=1}^{N}x_{i}}\right]^{3}\delta_{\iota,\iota'}\label{eq:RadiativeFinall}
\end{equation}
The factor in front of the delta we will call $a(J,\Lambda)$:
\begin{equation}
a(J,\Lambda):=\ln\Lambda\cdot\left(\frac{1}{J}\right)^{3}\cdot\frac{27}{32\pi}\left[\frac{1}{\left(1+\gamma^{2}\right)\sum_{i=1}^{N}x_{i}}\right]^{3}\label{eq:aJL}
\end{equation}
We will use it in the next subsection in the edge-renormalization problem.

Note that since $\Lambda$ is the maximum spin appearing in the bubble
causing infinity, whereas $J$ is the maximum spin appearing on the
external faces, they cannot be identified. To make the formula
\eref{eq:RadiativeFinall} correct we have to assume $1\ll J\ll\Lambda$.

\subsection{Application in edge renormalization}
\begin{figure}[ht!] 
	\centering
	\includegraphics[width=0.90\textwidth]{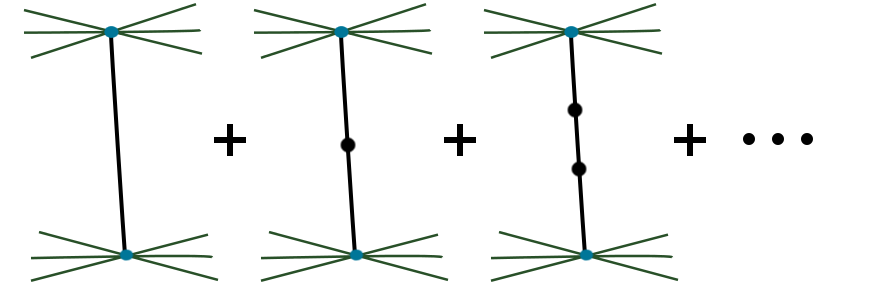}
	\caption{The foams $\kappa_0$, $\kappa_1$, $\kappa_2$, etc. form a geometric series, that can be summed to $\sum_{n=0}^\infty \kappa_n=\kappa_0\frac{1}{1-\T}$.}
	\label{fig:EdgeRen}
\end{figure}

Let us now consider a series of spin-foams with $\kappa_n$ such, 
that they differ only by a number of vertices on one selected edge $e$, 
i.e. $\kappa_n$ has $n$ vertices (see \fref{fig:EdgeRen}). Then the spin-foam 
operator related to each $\kappa_n$ is
\begin{equation}
W_{\kappa_n}=W_{\kappa_0}\cdot\T^n
\end{equation}
Obviously $\left\|\T\right\|<1$. Indeed, the factor 
$\alpha:=\left(\frac{1}{4\pi}\right)^{2}\left(\frac{6\pi}{\left(1+\gamma^{2}\right)\sum_{i=1}^{N}x_{i}}\right)^{\frac{3}{2}}<\frac{0.52}{\left(\left(1+\gamma^{2}\right)\sum_{i=1}^{N}x_{i}\right)^{3/2}}$ is less than $1$, because $1+\gamma^2\ge1$ and $\sum_{i=1}^{N}x_{i}\ge2$. Moreover by assumption $J\gg1$. Thus we can sum the leading orders in this series to
\begin{equation}
	W_\kappa^{\rm R}=\sum_{n=0}^\infty W_{\kappa_n}=W_{\kappa_0}\frac{1}{1-\alpha J^{-3/2}}
\end{equation}

\begin{figure}[ht!] 
	\centering
	\includegraphics[width=0.90\textwidth]{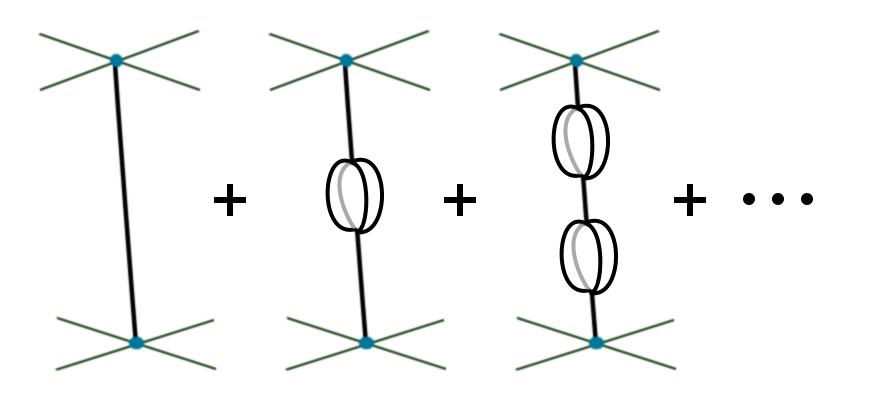}
	\caption{The foams with increasing number of ``melonic'' bubbles $\tilde\kappa_0$, $\tilde\kappa_1$, $\tilde\kappa_2$, etc. form a geometric series, that can be summed to $\sum_{n=0}^\infty \tilde\kappa_n=\tilde\kappa_0\frac{1}{1-\ln\Lambda\T^2}$.}
	\label{fig:BubbleRen}
\end{figure}

The same procedure can be done for a series of spin-foams $\tilde\kappa_N$ that differ by a number of bubbles on a 4-valent edge (see \fref{fig:BubbleRen}). Then, using \eref{eq:RadiativeFinall} we get
\begin{equation}\label{eq:W_Renormalized}
	W_{\tilde\kappa}^{\rm R,bubble}=\sum_{n=0}^\infty W_{\tilde\kappa_n}=W_{\tilde\kappa_0}\frac{1}{1-a(J,\Lambda)}
\end{equation}
The above formula is convergent iff $a(J,\Lambda)<1$.

Assume now for a moment, that the maximum spin $\Lambda$ is the inverse cosmological constant expressed in Plank units, i.e. that $\Lambda=10^{120}$ (such choice is justified in \cite{Aldo}). Then we get an upper bound:
\begin{equation}
a(J,\Lambda)<\frac{9.28}{J^3}
\end{equation}
thus for $J\ge2\frac12$ the sum \eref{eq:W_Renormalized} is convergent. Note, that all our approximations were made in large $J$ limit, so they do not apply for $J\le2$, and thus it is possible, that the sum \eref{eq:W_Renormalized} is convergent for all $J$s.

\section{Conclusions and further directions\label{sec:Conclusions}}

The Lorentzian Polyhedra Propagator is an important object in Spin-foam theory. It clarifies behaviour of the edge amplitude. It adds precision to calculations of Dipole Cosmology and ``melonic'' bubble divergence. It opens a path to study renormalization.

The tools developed in this study may be useful in direct computation of more complicated transitions amplitudes. The matrix elements of Lorentzian EPRL representation are key objects in Spin-foam calculations, and their properties were investigated here (mainly in the appendices).

A natural following step is to study subleading order of the $\T$ operator. We expect $\T$ can be approximated by an operator similar to a heat kernel and the subleading order would describe its spread.

Another further direction is to study in more detail the issue of renormalization in Spin-foam models. It contains both the low-$J$ behaviour of bubble divergence, more complicated bubbles (i.e. bubbles more than 4-valent edges) and higher order bubbles.

\ack{}{}

Firstly I would like to thank a lot Carlo Rovelli and Aldo Riello
for inspiring discussions and valuable remarks. Many thanks to Wojciech
Kami\'{n}ski, Marcin Kisielowski and J\k{e}drzej Swie\.{z}ewski for lots of
technical advice and to Jerzy Lewandowski for remarks on the draft
of the article. I also thank the participants of \emph{Emerging Fields
Initiative winter conference on canonical and covariant LQG},
for many interesting comments.

This work was supported by the Foundation for Polish Science International
PhD Projects Programme co-financed by the EU European Regional Development
Fund. 

\appendix

\section{Same calculations:\label{sec:AppHypergeometric}}

In the appendix let us present some technical details about calculations.

\subsection{Definitions\label{sub:AHDefinitions}}

Let us recall some definitions:

Firstly the Gauss hypergeometric function in its series and integral
representation:
\begin{equation}
\F\left(a,b;c;z\right):=\sum_{k=0}^{\infty}\frac{a^{\overline{k}}b^{\overline{k}}}{c^{\overline{k}}k!}z^{k}=\frac{\Gamma(c)}{\Gamma(b)\Gamma(c-b)}\int_{0}^{1}\dd t\, t^{b-1}\left(1-t\right)^{c-b-1}\left(1-zt\right)^{-a}\label{eq:Series2F1}
\end{equation}
where 
\begin{equation}
x^{\overline{k}}:=\prod_{n=0}^{k-1}(x+n)
\end{equation}
For $|z|<1$the series is convergent. Note that the if all arguments
$a,b,c>0$ and $0\le z<1$, then the series is sum of real positive
numbers, thus in this case $\F(a,b;c;z)\ge0$.

The function $f_{m}^{(j)}(\eta)$:
\begin{equation}
f_{m}^{(j)}(\eta)=e^{-(j+1+m-\im\gamma j)\eta}\ _{2}F_{1}\left(j+1-\im j\gamma,\ j+1+m;\ 2j+2;\ 1-e^{-2\eta}\right)
\end{equation}

Let us define another function, $g_{m}^{(j)}(\eta)$, by setting $\gamma=0$
in the above formula:
\begin{equation}
g_{m}^{(j)}(\eta)=e^{-(j+1+m)\eta}\ _{2}F_{1}\left(j+1,\ j+1+m;\ 2j+2;\ 1-e^{-2\eta}\right)
\end{equation}
One can easily see, that $g_{m}^{(j)}(\eta)\ge0$. Indeed, all the
parameters of $\F$function are positive.

In this section it is convenient to consider $f_{m}^{(j)}$ and $g_{m}^{(j)}$
as functions of $z(\eta)$:
\begin{equation}
f_{m}^{(j)}(\eta)=(1-z)^{\frac{(j+1+m-\im\gamma j)}{2}}\ _{2}F_{1}\left(j+1-\im j\gamma,\ j+1+m;\ 2j+2;\ z\right)
\end{equation}

We will also use a following shortcut notation:
\begin{equation}
J:=j+1\ \ \ \ G:=\gamma j\label{eq:JG_def}
\end{equation}
\begin{equation}
a:=J-\im G\ \ \ \ b:=J+m\ \ \ \ c:=2J\ \ \ \ \delta:=J+m-\im G\label{eq:abc_def}
\end{equation}
thus
\begin{equation}
\fjmn=\left(1-z\right)^{\frac{\delta}{2}}\FF abcz
\end{equation}
in case of $g_{m}^{(j)}\left(\eta\right)$ it is convenient to use
also $\tilde{a}:=J$, so
\begin{equation}
g_{m}^{(j)}\left(\eta\right)=\left(1-z\right)^{\frac{b}{2}}\FF{\tilde{a}}bcz
\end{equation}

\subsection{Some basic properties of $f_{m}^{(j)}(\eta)$ function\label{sub:AHBasic}}

\subsubsection{Mirror symmetry: $f_{m}^{(j)}(z)=\overline{f_{-m}^{(j)}(z)}$\label{sub:AHBasicMirror}}

Let us use the symmetry of $\F$ function:
\begin{equation}
\F(a,b;c;z)=(1-z)^{c-a-b}\F(c-a,c-b;c;z)
\end{equation}
and the mirror symmetry$\F(\overline{a},\overline{b};\overline{c};\overline{z})=\overline{\F(a,b;c;z)}$
to relate $f_{m}^{(j)}(z)$ with $f_{-m}^{(j)}(z)$. In our case $c-a=J+\im G=\overline{a}$
and $c-b=J-m$, and $c-a-b=-m+\im G$, thus for $0\le z<1$ we have
\begin{eqnarray}
f_{m}^{(j)}(z) & = & (1-z)^{\frac{\delta}{2}}\ _{2}F_{1}\left(J-\im G,\ J+m;\ 2J;\ z\right)\nonumber \\
\  & = & (1-z)^{\frac{J+m-\im G}{2}}(1-z)^{-m-\im G}\ _{2}F_{1}\left(J+\im G,\ J-m;\ 2J;\ z\right)\nonumber \\
\  & = & (1-z)^{\frac{J-m+\im G}{2}}\ _{2}F_{1}\left(J+\im G,\ J-m;\ 2J;\ z\right)\\
\  & = & \overline{(1-z)^{\frac{J-m-\im G}{2}}\ _{2}F_{1}\left(J-\im G,\ J-m;\ 2J;\ z\right)}=\overline{f_{-m}^{(j)}(z)}\nonumber 
\end{eqnarray}

Obviously the same calculation show that $g_{m}^{(j)}(z)=g_{-m}^{(j)}(z)$.

\subsubsection{First derivative of $f_{m}^{(j)}\left(z(\eta)\right)$\label{sub:AHBasicFirst}}

Let us now calculate the first derivative of $f_{m}^{(j)}$ and $g_{m}^{(j)}$
with respect to $\eta$. Using $\partial_{z}\F(a,b;c;z)=\frac{ab}{c}\F(a+1,b+1;c+1,z)$
we get: 
\begin{eqnarray}
\frac{\dd f_{m}^{(j)}\left(z(\eta)\right)}{\dd\eta} & = & -\delta e^{-\delta\eta}\F(a,b;c;z)\\
\  & \  & +e^{-\delta\eta}\frac{\dd z(\eta)}{\dd\eta}\frac{ab}{c}\F(a+1,b+1;c+1;z)\nonumber 
\end{eqnarray}
using \eref{eq:abc_def} and $z=1-e^{-2\eta}$ we get $\frac{\dd z}{\dd\eta}=2e^{-2\eta}=2(1-z)$,
$\frac{ab}{c}=\frac{(J-\im G)(J+m)}{2J}=\frac{\delta}{2}-\frac{\im Gm}{2J}$
so
\begin{eqnarray}
\frac{\dd f_{m}^{(j)}\left(z(\eta)\right)}{\dd\eta} & = & -\delta e^{-\delta\eta}\FF abcz\label{eq:dfjmn}\\
\  & \  & +e^{-(\delta+2)\eta}\left(\delta-\frac{\im Gm}{J}\right)\FF{a+1}{b+1}{c+1}z\nonumber 
\end{eqnarray}

At $\eta=0$ we use $z(0)=0$ and $\FF abc0=1$ to obtain
\begin{equation}
\detazero{\fjmn}=-\delta+\delta-\frac{\im Gm}{J}=-\frac{\im Gm}{J}=-\im\frac{\gamma jm}{j+1}=\underline{-\im m\left(1+O\left(j^{-1}\right)\right)}\label{eq:dfjmn0}
\end{equation}

To find the derivative of $g_{m}^{(j)}\left(\eta\right)$ we just
set $G=0$ in the above formulae and get
\begin{eqnarray}
\frac{\dd g_{m}^{(j)}\left(z(\eta)\right)}{\dd\eta} & = & -\delta e^{-\delta\eta}\FF abcz+\delta e^{-(\delta+2)\eta}\FF{a+1}{b+1}{c+1}z\nonumber \\
\  & = & -\delta e^{-\delta\eta}\left[\FF abcz-\left(1-z\right)\FF{a+1}{b+1}{c+1}z\right]
\end{eqnarray}
We can simplify the hypergeometric term using the series expansion:
\begin{eqnarray*}
\FF abcz-\left(1-z\right)\FF{a+1}{b+1}{c+1}z & = & \ 
\end{eqnarray*}
\begin{eqnarray}
\  & = & 1+\sum_{k=1}^{\infty}\frac{a^{\overline{k}}b^{\overline{k}}}{c^{\overline{k}}k!}z^{k}-1-\sum_{k=1}^{\infty}\frac{(a+1)^{\overline{k}}(b+1)^{\overline{k}}}{(c+1)^{\overline{k}}k!}z^{k}+\sum_{k=1}^{\infty}\frac{(a+1)^{\overline{k-1}}(b+1)^{\overline{k-1}}}{(c+1)^{\overline{k-1}}(k-1)!}z^{k}\nonumber \\
\  & = & \sum_{k=1}^{\infty}\frac{a^{\overline{k}}b^{\overline{k}}}{c^{\overline{k}}k!}z^{k}\left[1-\frac{(a+k)(b+k)c}{ab(c+k)}+\frac{ck}{ab}\right]\nonumber \\
\  & = & \sum_{k=1}^{\infty}\frac{a^{\overline{k}}b^{\overline{k}}}{c^{\overline{k}}k!}z^{k}\frac{abc+abk-abc-akc-bkc-ck^{2}+c^{2}k+ck^{2}}{ab(c+k)}\\
\  & = & \sum_{k=1}^{\infty}\frac{a^{\overline{k}}b^{\overline{k}}}{c^{\overline{k}}k!}z^{k}\frac{(c-a)(c-b)k}{ab(c+k)}\nonumber \\
\  & = & z\frac{(c-a)(c-b)}{c(c+1)}\sum_{k=1}^{\infty}\frac{(a+1)^{\overline{k-1}}(b+1)^{\overline{k-1}}}{(c+2)^{\overline{k-1}}(k-1)!}z^{k-1}\nonumber \\
\  & = & z\frac{(c-a)(c-b)}{c(c+1)}\sum_{k=1}^{\infty}\frac{(a+1)^{\overline{k}}(b+1)^{\overline{k}}}{(c+2)^{\overline{k}}k!}z^{k}=\nonumber \\
\  & = & z\frac{\left(c-a\right)\left(c-b\right)}{c\left(c+1\right)}\FF{a+1}{b+1}{c+2}z\nonumber 
\end{eqnarray}
so
\begin{equation}
\frac{\dd g_{m}^{(j)}\left(z(\eta)\right)}{\dd\eta}=-\delta e^{-\delta\eta}z(\eta)\frac{\left(c-a\right)\left(c-b\right)}{c\left(c+1\right)}\FF{a+1}{b+1}{c+2}{z(\eta)}
\end{equation}
what we will need in \eref{sub:AHEstymations}.

\subsubsection{Second derivative of $f_{m}^{(j)}\left(z(\eta)\right)$ and $\ln\left[f_{m}^{(j)}\left(z(\eta)\right)\right]$
at $\eta=0$\label{sub:AHBasicSecond}}

In \sref{sub:ISecondDerivative} we need to know $\frac{\dd^{2}}{\dd\eta^{2}}\ln\left[f_{m}^{(j)}\left(z(\eta)\right)\right]$
at $\eta=0$.

Let us first calculate $\frac{\dd^{2}f_{m}^{(j)}\left(z(\eta)\right)}{\dd\eta^{2}}$.
By differentiating the equation \eref{eq:dfjmn} and recalling, that
$\left(\delta-\frac{\im Gm}{J}\right)=2\frac{ab}{c}$we get
\begin{eqnarray}
\detadwa{\fjm\left(z(\eta)\right)} & = & -\delta\left[-\delta e^{-\delta\eta}\FF abc{z(\eta)}+2\frac{ab}{c}e^{-(\delta+2)\eta}\FF{a+1}{b+1}{c+1}{z(\eta)}\right]\nonumber \\
\  & \  & -\left(\delta+2\right)2\frac{ab}{c}e^{-(\delta+2)\eta}\FF{a+1}{b+1}{c+1}{z(\eta)}\\
\  & \  & +2\frac{ab}{c}\frac{\left(a+1\right)\left(b+1\right)}{\left(c+1\right)}\deta{z(\eta)}e^{-(\delta+2)\eta}\FF{a+2}{b+2}{c+2}{z(\eta)}\nonumber 
\end{eqnarray}
Fortunately we do not need to know the second derivative for all $\eta$,
it is enough to find its value at $\eta=0$, where $z(0)=0$ and $\F(a,b;c;0)=1$,
so the formula simplifies:
\begin{eqnarray}
\detadwazero{\fjm\left(z(\eta)\right)} & = & -\delta\left[-\delta+2\frac{ab}{c}\right]-\left(\delta+2\right)2\frac{ab}{c}+4\frac{ab}{c}\frac{\left(a+1\right)\left(b+1\right)}{\left(c+1\right)}
\end{eqnarray}
Now using \eref{eq:abc_def} one get
\begin{eqnarray}
\detadwazero{\fjm\left(z(\eta)\right)} & = & 4\left[\frac{\delta^{2}}{4}-\frac{ab\delta}{c}-\frac{ab}{c}+\frac{ab}{c}\frac{\left(a+1\right)\left(b+1\right)}{\left(c+1\right)}\right]
\end{eqnarray}

We are interested in the second differential of $\ln\left[\fjm\left(z(\eta)\right)\right]$.
Using the identity 
\begin{equation}
\frac{\dd^{2}\ln f(x)}{\dd x^{2}}=\frac{f''(x)f(x)-\left[f'(x)\right]^{2}}{\left[f(x)\right]^{2}}
\end{equation}
and the fact, that $\fjm(z(0))=1$ we have
\begin{equation}
\detadwazero{\ln\left[\fjm\left(z(\eta)\right)\right]}=\detadwazero{\fjm\left(z(\eta)\right)}-\left(\detazero{\fjm\left(z(\eta)\right)}\right)^{2}
\end{equation}
putting here the expressions \textbackslash{}reef and \textbackslash{}reef
we get
\begin{eqnarray}
\detadwazero{\ln\left[\fjm\left(z(\eta)\right)\right]} & = & 4\left[\frac{\delta^{2}}{4}-\frac{ab\delta}{c}-\frac{ab}{c}+\frac{ab}{c}\frac{\left(a+1\right)\left(b+1\right)}{\left(c+1\right)}\right]-\left[-\delta+2\frac{ab}{c}\right]^{2}\nonumber \\
\  & = & \delta^{2}-4\frac{ab\delta}{c}-4\frac{ab}{c}+4\frac{ab}{c}\frac{\left(a+1\right)\left(b+1\right)}{\left(c+1\right)}-\delta^{2}+4\frac{ab\delta}{c}-4\left(\frac{ab}{c}\right)^{2}\nonumber \\
\  & = & 4\frac{ab}{c}\left[-1-\frac{ab}{c}+\frac{\left(a+1\right)\left(b+1\right)}{\left(c+1\right)}\right]\\
\  & = & 4\frac{ab}{c}\cdot\frac{-c^{2}-c-abc-ab+abc+ac+bc+c}{c\left(c+1\right)}\nonumber \\
\  & = & -4\frac{ab}{c}\frac{\left(c-a\right)\left(c-b\right)}{c\left(c+1\right)}\nonumber \\
\  & = & -4\frac{a\left(c-a\right)\cdot b\left(c-b\right)}{c^{2}\left(c+1\right)}\nonumber 
\end{eqnarray}

Now using \eref{eq:abc_def} we have
\begin{eqnarray}
\detadwazero{\ln\left[\fjm\left(z(\eta)\right)\right]} & = & -4\frac{\left(J^{2}+G^{2}\right)\left(J^{2}-m^{2}\right)}{4J^{2}\left(2J+1\right)}=-\left(1+\frac{G^{2}}{J^{2}}\right)\frac{J^{2}-m^{2}}{\left(2J+1\right)}
\end{eqnarray}
and using \eref{eq:JG_def}
\begin{eqnarray}
\detadwazero{\ln\left[\fjm\left(z(\eta)\right)\right]} & = & -\left(1+\left(\frac{\gamma j}{j+1}\right)^{2}\right)\frac{\left(j+1\right)^{2}-m^{2}}{\left(2j+3\right)}
\end{eqnarray}
Since $\frac{\gamma j}{j+1}=\gamma\left(1+O\left(j^{-1}\right)\right)$,
the leading order of $\detadwazero{\ln\left[\fjm\left(z(\eta)\right)\right]}$
is
\begin{equation}
\detadwazero{\ln\left[\fjm\left(z(\eta)\right)\right]}=-\left(1+\gamma^{2}\right)\frac{\left(j+1\right)^{2}-m^{2}}{\left(2j+3\right)}\left(1+O\left(j^{-1}\right)\right)
\end{equation}

The term dependent on $m$ can be further simplified - see \ref{sec:AppSpin}.

\subsection{Estimations\label{sub:AHEstymations}}

It is easy to see, that $|\left|f_{m}^{(j)}(\eta)\right|\le g_{m}^{(j)}(\eta)$.
Indeed, taking the integral representation of $\F$ we see, that
\begin{eqnarray}
\left|f_{m}^{(j)}\left(z(\eta)\right)\right| & = & \left|e^{-(j+1+m-\im\gamma j)\eta}\ \frac{(2j+1)!}{(j+m)!(j-m)!}\int_{0}^{1}\dd t\,\frac{t^{j+m}\left(1-t\right)^{j-m}}{\left[1-\left(1-e^{-2\eta}\right)t\right]^{(j+1)-\im\gamma j}}\right|\nonumber \\
\  & = & e^{-(j+1+m)\eta}\ \frac{(2j+1)!}{(j+m)!(j-m)!}\left|\int_{0}^{1}\dd t\,\frac{t^{j+m}\left(1-t\right)^{j-m}}{\left[1-\left(1-e^{-2\eta}\right)t\right]^{(j+1)-\im\gamma j}}\right|\nonumber \\
\  & \le & e^{-(j+1+m)\eta}\ \frac{(2j+1)!}{(j+m)!(j-m)!}\ \int_{0}^{1}\dd t\,\left|\frac{t^{j+m}\left(1-t\right)^{j-m}}{\left[1-\left(1-e^{-2\eta}\right)t\right]^{(j+1)-\im\gamma j}}\right|\\
\  & = & e^{-(j+1+m)\eta}\ \frac{(2j+1)!}{(j+m)!(j-m)!}\ \int_{0}^{1}\dd t\, t^{j+m}\left(1-t\right)^{j-m}\left[1-\left(1-e^{-2\eta}\right)t\right]^{-(j+1)}\nonumber \\
\  & = & e^{-(j+1+m)\eta}\ \F\left(j+1,\ j+1;\ 2j+2;\ 1-e^{-2\eta}\right)=g_{m}^{(j)}\left(z(\eta)\right)\nonumber 
\end{eqnarray}
Let us now estimate $g_{m}^{(j)}(\eta)$ from above and from below.

To estimate $g_{m}^{(j)}$ we will estimate it's derivative. We will
consider bounds of $\frac{\frac{\dd g_{m}^{(j)}}{\dd\eta}}{g_{m}^{(j)}}$:
\begin{equation}
\frac{\frac{\dd g_{m}^{(j)}}{\dd\eta}\left(z(\eta)\right)}{g_{m}^{(j)}}=-\frac{(J+m)(J-m)}{2(2J+1)}\ \frac{z\F(J+1,\ J+m+1;\ 2(J+1);\ z)}{\F(J,\ J+m;\ 2J;\ z)}\label{eq:dg_estimation}
\end{equation}
To simplify the formulae let us introduce a constant $\omega:=\frac{(J+m)(J-m)}{2(2J+1)}=\frac{b\left(c-b\right)}{2\left(c+1\right)}$.

\subsubsection{Estimation of $g_{m}^{(j)}(z)$ from above\label{sub:AHEstymationAbove}}

One can easily see, that for nonnegative $a,b,c$, for $0\le z<1$
and for $c=2a$ and $b\le c$ the following inequality holds
\begin{equation}
\F(a,b;c;z)\le\F(a+1,b+1;c+2;z)
\end{equation}
 Indeed, using the series representation we can compare each term,
obtaining:
\begin{eqnarray}
1 & = & 1\ \ \ (k=0)\nonumber \\
\frac{ab}{2a} & < & \frac{(a+1)(b+1)}{2a+2}\ \ \ (k=1)\\
\frac{ab}{(2a)(2a+1)} & < & \frac{(a+k)(b+k)}{(2a+k)(2a+k+1)}\ \ \ (k\ge2)\nonumber 
\end{eqnarray}
where the last inequality is equivalent to
\begin{equation}
0<k^{2}a\left(4a+2-b\right)+k\left(4a^{2}+2a^{2}+ab\right)
\end{equation}
which is satisfied for our assumptions. Thus $\frac{z\F(J+1,\ J+m+1;\ 2J+2;\ z)}{\F(J,\ J+m;\ 2J;\ z)}\ge z$,
so
\begin{equation}
\frac{\frac{\dd g_{m}^{(j)}}{\dd\eta}\left(z(\eta)\right)}{g_{m}^{(j)}}\le-\frac{(J+m)(J-m)}{2(2J+1)}z(\eta)=-\omega\left(1-e^{-2\eta}\right)
\end{equation}
Integrating the above equations one gets
\begin{equation}
\left.\ln\left(g_{m}^{(j)}\right)\right|_{0}^{\eta}\le\frac{\omega}{2}-\omega\eta-\frac{\omega}{2}e^{-2\eta}
\end{equation}
so we have the upper bound:
\begin{equation}
g_{m}^{(j)}\left(z(\eta)\right)\le\left[e^{1-e^{-2\eta}-2\eta}\right]^{\frac{(J+m)(J-m)}{4(2J+1)}}
\end{equation}
and replacing $J$ by $j+1$ (see \eref{eq:JG_def})
\begin{equation}
\underline{g_{m}^{(j)}\left(z(\eta)\right)\le\left[e^{1-e^{-2\eta}-2\eta}\right]^{\frac{(j+1+m)(j+1-m)}{4(2j+3)}}}
\end{equation}

\subsubsection{Estimation of $g_{m}^{(j)}(z)$ from below\label{sub:AHEstymationBelow}}

To find the lower bound of \eref{eq:dg_estimation} let us show,
that under even weaker assumptions (i.e. $a,b,c\ge0$ and $0\le z<1$)
the following is true:
\begin{equation}
\exists_{\alpha>0}\alpha\F(a,b;c;z)\ge z\F(a+1,b+1;c+2;z)
\end{equation}
Let us introduce auxiliary quantity $A_{\alpha}:=\alpha\F(a,b;c;z)-z\F(a+1,b+1;c+2;z)$
and use the series expansion of the hypergeometric functions:
\begin{eqnarray}
A_{\alpha} & = & \alpha+\alpha\sum_{k=1}^{\infty}\frac{a^{\overline{k}}b^{\overline{k}}}{c^{\overline{k}}k!}z^{k}-\sum_{k=1}^{\infty}\frac{\left(a+1\right)^{\overline{k-1}}\left(b+1\right)^{\overline{k-1}}}{\left(c+2\right)^{\overline{k-1}}(k-1)!}z^{k}\nonumber \\
\  & = & \alpha+\sum_{k=1}^{\infty}\frac{a^{\overline{k}}b^{\overline{k}}}{c^{\overline{k}}k!}z^{k}\left(\alpha-\frac{c(c+1)k}{ab(c+k)}\right)\nonumber \\
\  & = & \alpha+\sum_{k=1}^{\infty}\frac{a^{\overline{k}}b^{\overline{k}}}{c^{\overline{k}}k!}z^{k}\left(\frac{\alpha abc+\alpha abk-c(c+1)k}{ab(c+k)}\right)\\
\  & = & \alpha+\alpha\sum_{k=1}^{\infty}\frac{a^{\overline{k}}b^{\overline{k}}}{\left(c+1\right)^{\overline{k}}k!}z^{k}+\frac{\left(\alpha ab-c(c+1)\right)}{c(c+1)}\sum_{k=1}^{\infty}\frac{\left(a+1\right)^{\overline{k-1}}\left(b+1\right)^{\overline{k-1}}}{\left(c+2\right)^{\overline{k-1}}(k-1)!}z^{k}\nonumber \\
\  & = & \alpha\F(a,b;c+1;z)+\frac{z}{c(c+1)}\left(\alpha ab-c(c+1)\right)\F(a+1,b+1;c+2;z)\nonumber 
\end{eqnarray}
For $\alpha\ge\alpha_{0}:=\frac{c(c+1)}{ab}$ all the elements of
the above formula are nonnegative, thus we have $\left.A_{\alpha}\ge0\right|_{\alpha\ge\alpha_{0}}$,
which proves the lemma. Thus
\begin{equation}
\frac{z\F(a+1,b+1;c+2;z)}{\F(a,b;c;z)}\le\alpha_{0}
\end{equation}
and so
\begin{equation}
\frac{\frac{\dd g_{m}^{(j)}}{\dd\eta}\left(z(\eta)\right)}{g_{m}^{(j)}}\ge-\omega\alpha_{0}=-\frac{b\left(c-b\right)}{2\left(c+1\right)}\frac{c\left(c+1\right)}{ab}=-\frac{c}{2a}\left(c-b\right)
\end{equation}
In our case $c=2a$, so
\begin{equation}
\frac{\frac{\dd g_{m}^{(j)}}{\dd\eta}\left(z(\eta)\right)}{g_{m}^{(j)}}\ge-\left(c-b\right)=-\left(j+1-m\right)
\end{equation}
Integrating above formula one gets
\begin{equation}
g_{m}^{(j)}\left(z(\eta)\right)\ge e^{-\eta(j+1-m)}
\end{equation}

So finally we have
\begin{equation}
\underline{e^{-\eta(j+1-m)}\le g_{m}^{(j)}\left(z(\eta)\right)\le\left[e^{1-e^{-2\eta}-2\eta}\right]^{\frac{(j+1+m)(j+1-m)}{4(2j+3)}}}
\end{equation}

\section{Squared magnetic momentum number \label{sec:AppSpin}}

Several times in the calculations above the squared magnetic number
$m^{2}$ appears. It seem to break $SU(2)$ invariance, however when
considering the invariant states $\ket{\iota}\in{\rm Inv}\left(\Hil_{j_{1}}\otimes\cdots\otimes\Hil_{j_{N}}\right)$,
one can express such components in terms of gauge invariant quantities,
what we will prove below.

At the beginning let us remind, that given an invariant state $\ket{\iota}$
we can decompose it in the magnetic momentum basis
\begin{equation}
\ket{\iota}=\sum_{m_{1},\ldots,m_{N}}\iota_{m_{1}\cdots m_{N}}\ket{m_{1}}_{j_{1}}\otimes\cdots\otimes\ket{m_{N}}_{j_{N}}=:\sum_{\vec{m}}\iota_{\vec{m}}\ket{\vec{m}}_{\vec{j}}
\end{equation}
note, that
\begin{equation}
\iota_{\vec{m}}:=\bra{\vec{m}}\ket{\iota}_{\vec{j}}\label{eq:iota_m}
\end{equation}

We will learn how to compute the formulae of the form
\begin{equation}
\sum_{\vec{m}}f\left(m_{a}^{2},j_{b}\right)\overline{\iota{}_{\vec{m}}}\iota'_{\vec{m}}
\end{equation}
for a real analytic function $f$.

\subsection{Single squared magnetic momentum number\label{sub:ASSingleSquaredMagneticNumber}}

First let us consider an expression $\sum_{\vec{m}}m_{i}^{2}\overline{\iota{}_{\vec{m}}}\iota'_{\vec{m}}$
for a single index $i$. By definition we have $m_{i}^{2}\ket{\vec{m}}_{\vec{j}}=\widehat{L_{z,(i)}}^{2}\ket{\vec{m}}_{\vec{j}}$.
Now using \eref{eq:iota_m} we get
\begin{eqnarray}
\sum_{\vec{m}}m_{i}^{2}\overline{\iota{}_{\vec{m}}}\iota'_{\vec{m}} & = & \sum_{\vec{m}}\bra{\iota}\ket{\vec{m}}_{\vec{j}}\bra{\vec{m}}\ket{\iota'}_{\vec{j}}m_{i}^{2}\nonumber \\
\  & = & \sum_{\vec{m}}\bra{\iota}m_{i}^{2}\ket{\vec{m}}_{\vec{j}}\bra{\vec{m}}\ket{\iota'}_{\vec{j}}\\
\  & = & \sum_{\vec{m}}\bra{\iota}\widehat{L_{z,(i)}}^{2}\ket{\vec{m}}_{\vec{j}}\bra{\vec{m}}\ket{\iota'}_{\vec{j}}\nonumber 
\end{eqnarray}
Now since $\sum_{\vec{m}}\ket{\vec{m}}\bra{\vec{m}}$ is the identity
operator, we have
\begin{equation}
\sum_{\vec{m}}m_{i}^{2}\overline{\iota{}_{\vec{m}}}\iota'_{\vec{m}}=\bra{\iota}\widehat{L_{z,(i)}}^{2}\ket{\iota'}
\end{equation}
Now thanks to $SU(2)$ invariance we have
\begin{eqnarray}
\bra{\iota}\widehat{L_{z,(i)}}^{2}\ket{\iota'} & = & \bra{\iota}\widehat{L_{x,(i)}}^{2}\ket{\iota'}=\bra{\iota}\widehat{L_{y,(i)}}^{2}\ket{\iota'}\nonumber \\
\  & = & \frac{1}{3}\bra{\iota}\left[\widehat{L_{x,(i)}}^{2}+\widehat{L_{y,(i)}}^{2}+\widehat{L_{z,(i)}}^{2}\right]\ket{\iota'}\label{eq:Lz2_as_L2}\\
\  & = & \frac{1}{3}\bra{\iota}\widehat{L_{(i)}^{2}}\ket{\iota'}\nonumber 
\end{eqnarray}
The operator $\widehat{L_{(i)}^{2}}$ is an invariant with eigenvalue
$j_{i}(j_{i}+1)$, thus after all we have 
\begin{equation}
\sum_{\vec{m}}m_{i}^{2}\overline{\iota{}_{\vec{m}}}\iota'_{\vec{m}}=\frac{j_{i}(j_{i}+1)}{3}\bra{\iota}\ket{\iota'}
\end{equation}

\subsection{Real function of $m_{i}^{2}$\label{sub:ASRealFunctionOfM2}}

Consider now a real analytic function $f\left(m_{i}^{2}\right)$ instead
of $m_{i}^{2}$, i.e. let us calculate $\sum_{\vec{m}}f\left(m_{i}^{2}\right)\overline{\iota{}_{\vec{m}}}\iota'_{\vec{m}}$.
We will start with a polynomial: $\left(m_{i}^{2}\right)^{n}$. Since
$\widehat{L_{z,(i)}}^{2}$ is a positive, selfadjoint operator, we
can repeat the above procedure and obtain
\begin{equation}
\sum_{\vec{m}}\left(m_{i}^{2}\right)^{n}\overline{\iota{}_{\vec{m}}}\iota'_{\vec{m}}=\bra{\iota}\left(\widehat{L_{z,(i)}}^{2}\right)^{n}\ket{\iota'}
\end{equation}
Now we can insert an orthonormal intertwiner basis between each two
$\widehat{L_{z,(i)}}^{2}$ operators:
\begin{equation}
\sum_{\vec{m}}\left(m_{i}^{2}\right)^{n}\overline{\iota{}_{\vec{m}}}\iota'_{\vec{m}}=\sum_{\iota_{1\cdots\iota_{n-1}}}\bra{\iota}\widehat{L_{z,(i)}}^{2}\ketbra{\iota_{1}}\cdots\ketbra{\iota_{n-1}}\widehat{L_{z,(i)}}^{2}\ket{\iota'}
\end{equation}
Each expression $\bra{\iota_{I}}\widehat{L_{z,(i)}}^{2}\ket{\iota_{I+1}}$
equals $\frac{1}{3}\bra{\iota_{I}}\widehat{L_{(i)}^{2}}\ket{\iota_{I+1}}$
(see \eref{eq:Lz2_as_L2}) giving the eigenvalue $\frac{j_{i}(j_{i}+1)}{3}$,
thus we have
\begin{equation}
\sum_{\vec{m}}\left(m_{i}^{2}\right)^{n}\overline{\iota{}_{\vec{m}}}\iota'_{\vec{m}}=\left[\frac{j_{i}(j_{i}+1)}{3}\right]^{n}\bra{\iota}\ket{\iota'}
\end{equation}

For a real analytic function $f$ more general than the polynomial
we expand $f$ in a power series and follow the above steps for each
power of $m_{i}^{2}$, obtaining
\begin{equation}
\sum_{\vec{m}}f\left(m_{i}^{2}\right)\overline{\iota{}_{\vec{m}}}\iota'_{\vec{m}}=f\left(\frac{j_{i}(j_{i}+1)}{3}\right)\bra{\iota}\ket{\iota'}
\end{equation}

\subsection{Function of many $m_{i}^{2}$s\label{sub:ASFunctionOfManyM2s}}

Consider now a real function $f\left(m_{1}^{2},\ldots,m_{N}^{2}\right)$.
Since the operators $\widehat{L_{z,(i)}}$ commute for different $i$,
for each term $m_{i}^{2}$ we can do the procedure of \ref{sub:ASRealFunctionOfM2}
separately, obtaining
\begin{equation}
\sum_{\vec{m}}f\left(m_{1}^{2},\ldots,m_{N}^{2}\right)\overline{\iota{}_{\vec{m}}}\iota'_{\vec{m}}=f\left(\frac{j_{1}(j_{1}+1)}{3},\ldots,\frac{j_{N}(j_{N}+1)}{3}\right)\bra{\iota}\ket{\iota'}
\end{equation}

\subsection{Function of many $m_{i}^{2}$s and many $j_{i}$s\label{sub:ASGeneralFunction}}

At least let us consider a real function $f\left(m_{1}^{2},\ldots,m_{N}^{2},j_{1},\ldots,j_{N}\right)=:f\left(\overrightarrow{m^{2}},\vec{j}\right)$.
Note, that given an expression $\sum_{\vec{m}}f\left(\overrightarrow{m^{2}},\vec{j}\right)\overline{\iota{}_{\vec{m}}}\iota'_{\vec{m}}$,
the $j_{i}$-dependent parts do not differ when $\vec{m}$ changes.
We can thus follow the procedure of \ref{sub:ASFunctionOfManyM2s},
treating all $j_{i}$-dependences as parameters of the function, and
obtain the result: 
\begin{equation}
\sum_{\vec{m}}f\left(\overrightarrow{m^{2}},\vec{j}\right)\overline{\iota{}_{\vec{m}}}\iota'_{\vec{m}}=f\left(\overrightarrow{\frac{j\left(j+1\right)}{3}},\vec{j}\right)\bra{\iota}\ket{\iota'}
\end{equation}

\subsection{Application in the text\label{sub:ASApplication}}

In the main text of the article the above lemma is used twice: when
the large $J$ behaviour of the function $\Phi_{\iota'}^{\iota}(\eta)$
is considered \sref{sub:IMaximumPoint}, and when the Hessian
matrix of $\phi_{\iota'}^{\iota}(\eta)$ is calculated \sref{sub:ISecondDerivative}.

In the first case we estimate $\left|\Phi_{\iota'}^{\iota}(\eta_{1})\right|=\left|\sum_{\vec{m}}\Phi_{\vec{m}}^{\vec{m}}(\eta_{1})\overline{\iota_{\vec{m}}}\iota'_{\vec{m}}\right|\le\left|\sum_{\vec{m}}\left|\Phi_{\vec{m}}^{\vec{m}}(\eta_{1})\right|\overline{\iota_{\vec{m}}}\iota'_{\vec{m}}\right|$,
knowing that $\left|\Phi_{\vec{m}}^{\vec{m}}(\eta_{1})\right|\le\left[C(\eta_{1})\right]^{\sum_{i=1}^{N}\frac{(j_{i}+1)^{2}-m_{i}^{2}}{4(2j_{i}+3)}}$,
where the factor $C(\eta_{1})$ does not depend on $\vec{m}$. We
have thus
\begin{equation}
\left|\Phi_{\iota'}^{\iota}(\eta_{1})\right|\le\left|\sum_{\vec{m}}\left[C(\eta_{1})\right]^{\sum_{i=1}^{N}\frac{(j_{i}+1)^{2}-m_{i}^{2}}{4(2j_{i}+3)}}\overline{\iota_{\vec{m}}}\iota'_{\vec{m}}\right|
\end{equation}
and using our lemma we can substitute each appearance of $m_{i}^{2}$
by $\frac{j_{i}\left(j_{i}+1\right)}{3}$, obtaining
\begin{equation}
\left|\Phi_{\iota'}^{\iota}(\eta_{1})\right|\le\left[C(\eta_{1})\right]^{\sum_{i=1}^{N}\frac{(j_{i}+1)^{2}-\frac{j_{i}\left(j_{i}+1\right)}{3}}{4(2j_{i}+3)}}\left|\bra{\iota}\ket{\iota'}\right|=\left[C(\eta_{1})\right]^{\frac{1}{12}\sum_{i=1}^{N}j_{i}+1}\left|\bra{\iota}\ket{\iota'}\right|\label{eq:AppEstimationOfPhi}
\end{equation}

In the second case we calculate $\detadwazero{\phi_{\iota'}^{\iota}}=\frac{1}{\bra{\iota}\ket{\iota'}}\sum_{\vec{m}}\detadwazero{\phi_{\vec{m}}}\overline{\iota_{\vec{m}}}\iota'_{\vec{m}}$,
knowing that $\detadwazero{\phi_{\vec{m}}}=-\frac{1+\gamma^{2}}{J}\sum_{i=1}^{N}\frac{\left(j_{i}+1\right)^{2}-m_{i}^{2}}{\left(2j_{i}+3\right)}$.
Again using our lemma allows to substitute each appearance of $m_{i}^{2}$
by $\frac{j_{i}\left(j_{i}+1\right)}{3}$, obtaining
\begin{equation}
\detadwazero{\phi_{\iota'}^{\iota}}=-\frac{1+\gamma^{2}}{J}\sum_{i=1}^{N}\frac{\left(j_{i}+1\right)^{2}-\frac{j_{i}\left(j_{i}+1\right)}{3}}{\left(2j_{i}+3\right)}\frac{\bra{\iota}\ket{\iota'}}{\bra{\iota}\ket{\iota'}}=-\frac{1+\gamma^{2}}{3J}\sum_{i=1}^{N}\left(j_{i}+1\right)\label{eq:AppDetaDwa_phi_zero}
\end{equation}

\section{Proofs of lemmas used in the text\label{sec:AppProofs}}

\subsection{The inequality $\forall_{\eta>0}C(\eta)<1$\label{sub:APInequality}}

Let us show the following inequality
\begin{equation}
\forall_{\eta\ge0}\ \ \ \left(e^{1-2\eta-e^{-2\eta}}\right)\le1
\end{equation}
where the equality holds only for $\eta=0$.

Obviously the case $\eta=0$ is satisfied:
\begin{equation}
e^{1-0-e^{0}}=e^{0}=1
\end{equation}

Since both sides of the inequality are nonnegative, we can take the
logarithm of the inequality
\begin{equation}
1-2\eta-e^{-2\eta}\overset{?}{<}0
\end{equation}
We can differentiate both sides of the inequality with respect to
$\eta$:
\begin{equation}
-2+2e^{-2\eta}<0
\end{equation}
which is obviously true for $\eta>0$. Thus for $\eta>0$
\begin{equation}
1-2\eta-e^{-2\eta}=1+\int_{0}^{\eta}\left(-2+2e^{-2\tilde{\eta}}\right)\dd\tilde{\eta}<1+\int_{0}^{\eta}0\dd\tilde{\eta}=1
\end{equation}
\emph{quod erat demonstrandum}.

\subsection{\label{sub:APEstimation}The estimation in \sref{eq:PhiEstimation}}

We want to estimate the integral 
\begin{equation}
I_{\eta_{\epsilon}}:=\int_{\eta_{\epsilon}}^{\infty}\dd\eta\left(\frac{\sinh\eta}{4\pi\eta}\right)^{2}\left(e^{1-2\eta-e^{-2\eta}}\right)^{\kappa}
\end{equation}
for $\kappa>3$ and $\eta_{\epsilon}>0$.

First let us note, that $\frac{1}{\eta}\le\frac{1}{\eta_{\epsilon}}$
for all $\eta$ in the integration range, thus
\begin{equation}
I_{\eta_{\epsilon}}\le\left(\frac{1}{4\pi\eta_{\epsilon}}\right)^{2}\int_{\eta_{\epsilon}}^{\infty}\dd\eta\left(\sinh\eta\right)^{2}\left(e^{1-2\eta-e^{-2\eta}}\right)^{\kappa}\le\cdots
\end{equation}
Now note, that $\left(\sinh\eta\right)^{2}\le e^{2\eta}$, so 
\begin{equation}
\cdots\le\left(\frac{1}{4\pi\eta_{\epsilon}}\right)^{2}\int_{\eta_{\epsilon}}^{\infty}\dd\eta\ e^{2\eta}\left(e^{1-2\eta-e^{-2\eta}}\right)^{\kappa}=\cdots
\end{equation}
Now let us change the integration variables to $x=e^{-2\eta}$ :
\begin{equation}
\cdots=\left(\frac{1}{4\pi\eta_{\epsilon}}\right)^{2}\frac{1}{2}\int_{0}^{e^{-2\eta_{\eta}}=:x_{\eta}}\dd x\frac{1}{x^{2}}\left(xe^{1-x}\right)^{\kappa}\le\cdots
\end{equation}
all the integration rage is the subset of $x\le1$, so $e^{1-x}\le e$,
thus we can remove the denominator:
\begin{equation}
\cdots\le\frac{1}{2}\left(\frac{e}{4\pi\eta_{\epsilon}}\right)^{2}\int_{0}^{x_{\epsilon}}\dd x\left(xe^{1-x}\right)^{\kappa-2}\le\cdots
\end{equation}
One can easily prove, that $\forall_{x\in[0,1]}x\le e^{x-1}$. Indeed,
for $x=1$ we have $1=1$, and the derivative of left-hand side is
$x'=1$ is bigger than the derivative of the right-hand side $\left(e^{x-1}\right)'=e^{x-1}$.
Knowing that, and since $\kappa>3$, we can estimate $\left(xe^{1-x}\right)^{\kappa-2}\le xe^{1-x}$,
and though
\begin{equation}
\cdots\le\frac{1}{2}\left(\frac{e}{4\pi\eta_{\epsilon}}\right)^{2}\int_{0}^{x_{\epsilon}}\dd x\ xe^{1-x}=\cdots
\end{equation}
which can be integrated by parts:
\begin{equation}
\cdots=\frac{1}{2}\left(\frac{e}{4\pi\eta_{\epsilon}}\right)^{2}\left[-\left(x+1\right)e^{1-x}\right]_{0}^{x_{\epsilon}}=\cdots
\end{equation}
Now putting back $x_{\epsilon}=e^{-2\eta_{\epsilon}}$ we get
\begin{equation}
I_{\eta_{\epsilon}}\le\frac{1}{2}\left(\frac{e}{4\pi\eta_{\epsilon}}\right)^{2}\left(e-e^{1-e^{-2\eta_{\epsilon}}}-e^{1-2\eta_{\epsilon}-e^{-2\eta_{\epsilon}}}\right)=\tilde{I}_{\eta_{\epsilon}}\label{eq:I_eta_estymation}
\end{equation}

\subsection{Proof of existence of $\eta_{\epsilon}$\label{sub:APExistence}}

Now let us proof, that for each $\epsilon>0$ there is $\eta_{\epsilon}$
such that $I_{\eta_{\epsilon}}\le\epsilon$. We will do it by showing
(using Darboux theorem), that the equation $\tilde{I}_{\eta_{\epsilon}}=\epsilon$
has a solution.

Let us check the limits of $\tilde{I}_{\eta_{\epsilon}}$. For $\eta_{\epsilon}\to0$
we have
\begin{equation}
\lim_{\eta_{\epsilon}\to0}\tilde{I}_{\eta_{\epsilon}}=\frac{1}{2}\left(\frac{e}{4\pi}\right)^{2}\left(e-2\right)\left(\frac{1}{\eta_{\epsilon}}\right)^{2}=+\infty
\end{equation}
For $\eta_{\epsilon}\to+\infty$ we have 
\begin{equation}
\lim_{\eta_{\epsilon}\to+\infty}\tilde{I}_{\eta_{\epsilon}}=\frac{1}{2}\left(\frac{e}{4\pi}\right)^{2}\left[\frac{1}{+\infty}\left(e-e^{1-0}-e^{-\infty}\right)\right]=0
\end{equation}

Thus $\tilde{I}_{\eta_{\epsilon}}$runs through all real positive
numbers, so for each $\epsilon>0$ there is $\eta_{\epsilon}$ being
the solution to the equation $\tilde{I}_{\eta_{\epsilon}}$, and since
$I_{\eta_{\epsilon}}\le\tilde{I}_{\eta_{\epsilon}}$ (what we have
shown in \eref{eq:I_eta_estymation}), it is obvious now, that
\begin{equation}
\forall_{\epsilon>0}\exists_{\eta_{\epsilon}}I_{\eta_{\epsilon}}\le\epsilon
\end{equation}

\section*{---------------------}

\end{document}